\documentclass[final,5p,times,twocolumn]{elsarticle}
\usepackage{epsfig}

\usepackage{amsmath,amssymb,amsfonts}
\usepackage{xfrac}
\usepackage{algorithmic}
\usepackage{graphicx}
\usepackage{textcomp}
\usepackage{xcolor}
\usepackage{color}
\usepackage{placeins}
\usepackage{float}
\usepackage{hyperref}
\usepackage{tabularx,colortbl}
\usepackage{subcaption}
\usepackage{comment}
\usepackage{gensymb}

\usepackage{multirow}
\usepackage{rotating}
\usepackage{makecell}
\usepackage{natbib}

\def\BibTeX{{\rm B\kern-.05em{\sc i\kern-.025em b}\kern-.08em
    T\kern-.1667em\lower.7ex\hbox{E}\kern-.125emX}}
\journal{Acta Astronautica}

\begin{document}

\begin{frontmatter}



\title{Wireless Power Transfer in Space using Flexible, Lightweight, Coherent Arrays}

\author[eca,caltech]{Alex Ayling}
\author[eca,caltech]{Austin Fikes}
\author[eca,caltech]{Oren S. Mizrahi\corref{cor1}}
\ead{orensimonmizrahi@gmail.com}
\author[eca,caltech]{Ailec Wu}
\author[caltech]{Raha Riazati}
\author[caltech]{Jesse Brunet}
\author[guru]{Behrooz Abiri}
\author[indie]{Florian Bohn}
\author[bgu]{Matan Gal-Katziri}
\author[caltech]{Mohammed Reza M. Hashemi}
\author[jpl]{Sharmila Padmanabhan}
\author[aws]{Damon Russell}
\author[caltech]{Ali Hajimiri}

\affiliation[eca]{organization={{These authors contributed equally and are listed alphabetically by last name.}}}
\affiliation[caltech]{organization={Department of Electrical Engineering, California Institute of Technology (Caltech)},
            city={Pasadena},
            state={CA},
            postcode={91125}, 
            country={U.S.A.}
}
\affiliation[guru]{organization={Formerly with Caltech; currently with GuRu Wireless, Inc.},
            city={Pasadena},
            state={CA},
            postcode={91101}, 
            country={U.S.A.}
}
\affiliation[indie]{organization={Formerly with Caltech; currently with indie Semiconductor, Inc.},
            city={Aliso Viejo},
            state={CA},
            postcode={92656}, 
            country={U.S.A.}
}
\affiliation[bgu]{organization={Formerly with Caltech; currently with the Department of Electrical Engineering, Ben-Gurion University},
            city={Beer Sheva},
            country={Israel}
}
\affiliation[jpl]{organization={Caltech - Jet Propulsion Laboratory},
            city={Pasadena},
            state={CA},
            postcode={91011}, 
            country={U.S.A.}
}
\affiliation[aws]{organization={Formerly with Caltech - Jet Propulsion Laboratory; currently with Amazon Web Services},
            city={Pasadena},
            state={CA},
            postcode={91125}, 
            country={U.S.A.}
}

\cortext[cor1]{Corresponding author: Oren S. Mizrahi}

\begin{abstract}
Space solar power (SSP), envisioned for decades as a solution for continuous, stable, and dynamically dispatchable clean energy, has seen tremendous interest and a number of experimental demonstrations in the last few years. A practical implementation has been elusive to date, owing to the high launch costs associated with heavy, rigid photovoltaic (PV) and wireless power transfer (WPT) arrays. Lightweight and flexible solutions for WPT have been demonstrated terrestrially but, to date, have not been deployed and tested in space. In this paper, we present an experimental space demonstration of a lightweight, flexible WPT array powered by custom radio frequency integrated circuits (RFICs). The transmit arrays, receive arrays, and the rest of the system were operated and tested for eight months in Low Earth Orbit (LEO). Results from these experiments, including pointing of the array's beam to Earth and its detection by a ground station, are presented and discussed in detail. Observations and results from this mission uncover existing strengths and weaknesses that inform future steps toward realizing SSP.
\end{abstract}




\end{frontmatter}

\section{Introduction}
Wireless power transfer (WPT) at distance could transform the nature of energy distribution and access, both on Earth and in space. This has wide implications for everything from consumer electronic devices to large scale energy generation and distribution. It can make up for gaps due to the volatile nature of clean energy sources like wind and solar, expand access to locations where physical connections are non-existent, and enable entirely new applications. This can be achieved using phased arrays consisting of large numbers of small sources working coherently to focus transmitted energy on one or more receivers. By adjusting the relative timing of the individual elements, energy can be quickly dispatched to numerous select locations simultaneously without physically moving the transmitter. 

\begin{figure}
    \centering
    \includegraphics[width=\linewidth]{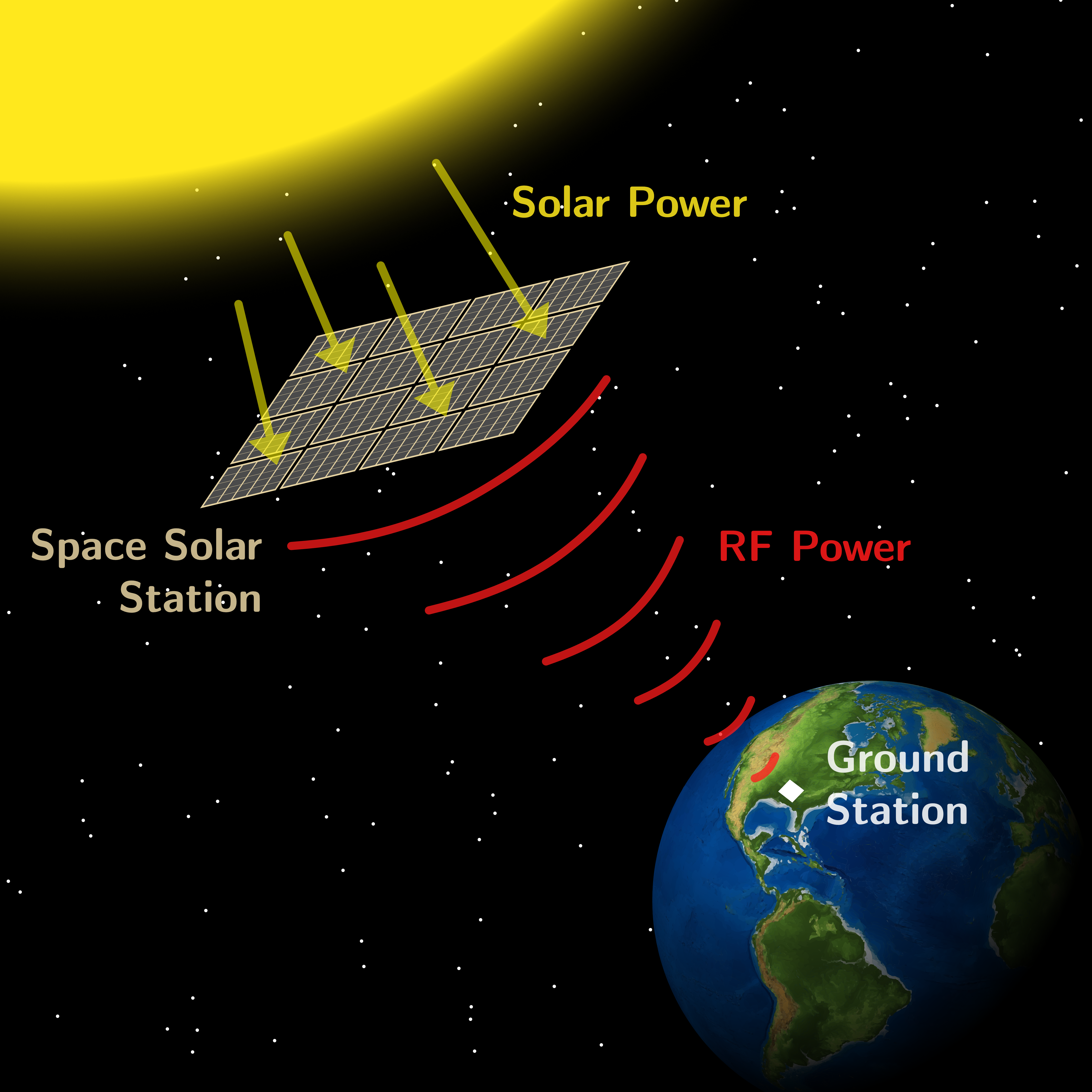}
    \caption{The Caltech SSP concept \cite{fikes2022caltech}: A PV-WPT array converts solar energy to DC power and DC power to RF power which is then collected on Earth by a ground station.}
    \label{fig:ssppConcept}
\end{figure}

The operational flexibility of WPT introduces a myriad of possibilities. While this is true for terrestrial WPT, it is especially relevant in space where physical connections between spacecraft, space stations, and ground stations are impractical if not impossible. In particular, WPT in space enables space solar power (SSP), a concept first conceived by Isaac Asimov in his short story, ``Reason" \cite{Asimov}, where solar energy is converted to microwaves in space and transmitted to Earth. Caltech's vision for SSP (Fig. \ref{fig:ssppConcept}) comprises the deployment of massive kilometer-scale arrays of photovoltaic (PV) cells in space \cite{abiri_2022_lightweight}. The solar power collected by the solar cells is first converted to DC power by the photovoltaics and then locally to RF power. Using phased arrays, the RF power is then wirelessly transmitted to destinations on Earth or, as envisioned by Azimov, to other planets inhabited by humans. Although light of any wavelength may be used to transfer energy wirelessly (e.g., lasers), RF/microwave wavelengths in certain frequency ranges are particularly well-suited for SSP due to lower atmospheric attenuation \cite{JAFFE_Paper}, permitting operation in all weather conditions. Also, the larger wavelengths in the microwave band (relative to lasers) combined with the physical diffraction limit can be used to guarantee that the power intensity on Earth will not exceed a predetermined maximum, alleviating safety concerns. A sufficiently large array will focus close to 90\% of its power in the main focal point at the receiver, with close to 100\% achievable by tapering the power towards the edges of the array \cite{JAFFE_Paper}. At the destination, the RF power is received and converted to low frequency (AC or DC) power using rectifying antenna (rectenna) arrays. Rectenna arrays for RF-DC conversion are well-studied and offer high conversion efficiency, with some systems demonstrating efficiencies as high as 70-80\% \cite{S_Band_High_Eff_Rectenna_Arr}. 

Unlike solar, wind, geothermal, or hydropower, SSP has the unique advantage of providing continuous, stable, and predictable renewable energy \cite{marshall2022dynamics} without imposing storage demands. Moreover, unlike nuclear, electronic architectures also enable dynamically dispatching power to nearly any desired location on Earth, including locations where the capacity or the infrastructure to satisfy the demand for electricity may not be available.

Significant work has been done on SSP over the last 50+ years, as meticulously documented in \cite{JAFFE_Paper, JOM_article}. Early work in the 1960s and 1970s by private and government institutions \cite{Raytheon_Helicopter, GLASER_Article, Raytheon_WilliamBrown_SSPP, Brown_IEEE, JPL_Lab_WPT, JPL_Goldstone_WPT} culminated in NASA's system report in 1979 \cite{NASA1979}. Progress continued with Japan's ISY-METS mission in 1993, which demonstrated sub-orbital WPT \cite{JAXA_High_Altitude_WPT}. New SSP concepts were developed by NASA \cite{MANKINS2002369}, ESA \cite{SEBOLDT2001785}, and JAXA \cite{MORI2006132, SASAKI2007153} in the mid 1990s and early 2000s. SSP concept proposals are enjoying an accelerating push towards a more renewable future \cite{CASH2019170, YANG201651, LI2017182}.

Despite the potential benefits of SSP and the large body of existing research, the enormous launch costs associated with deploying a massive array in high-altitude orbits have, to date, prohibited practical implementations of any SSP system. To reduce the overall system mass and cost, lightweight, flexible phased arrays powered by thin-film photovoltaic cells on one or both sides can be used. Radio frequency integrated circuits (RFICs) further reduce mass and size by combining onto a single millimeter-scale chip all the necessary digital and analog circuitry for DC to RF power conversion and focusing. These RFICs combined with the thin, lightweight PV and flexible, lightweight RF arrays reduce launch costs and help bring SSP closer to fruition \cite{mizrahi_TEA}.

Recently, such flexible, lightweight, coherent arrays powered by custom RFICs have been developed and tested on Earth \cite{hashemi2019flexible, kelzenberg_ultralight, gal2020scalable, gal2022flexible, fikes2022frontiers, FIKES_FRAMEWORK_MTT, MIZRAHI_FLEXIBLE_RECONSTRUCTION}. However, to the best of our knowledge, a dedicated free-space WPT system using this or any other technology has not been deployed and demonstrated in space.

MAPLE ({\bf M}icrowave {\bf A}rray {\bf P}ower Transfer {\bf L}EO {\bf E}xperiment) is an experiment to verify the ability of such arrays to transmit energy in space and to evaluate their strengths and weaknesses. Mission results will inform the next phase(s) of development of lightweight, flexible, coherent transmission arrays for energy transfer.

The remainder of the paper is organized as follows: Section 2 covers system design, including a discussion of the main hardware components, software design, mechanical and thermal considerations, and space qualification testing. Section 3 covers results from in-orbit experiments, including demonstrations of wireless power transmission and reception at multiple targets, focusing algorithm operation and performance, and a comparison to performance on Earth. Section 4 covers the system design and results for the ``beam-to-Earth" experiment, during which power generated by MAPLE was successively focused at and detected on Earth. Section 5 examines anomalous results and presents a hypothesis about the likely causes. Section 6 concludes the paper and briefly discusses future work.

\begin{figure}[t]
\centering
\begin{subfigure}[b]{0.48\textwidth}
   \includegraphics[width=\linewidth]{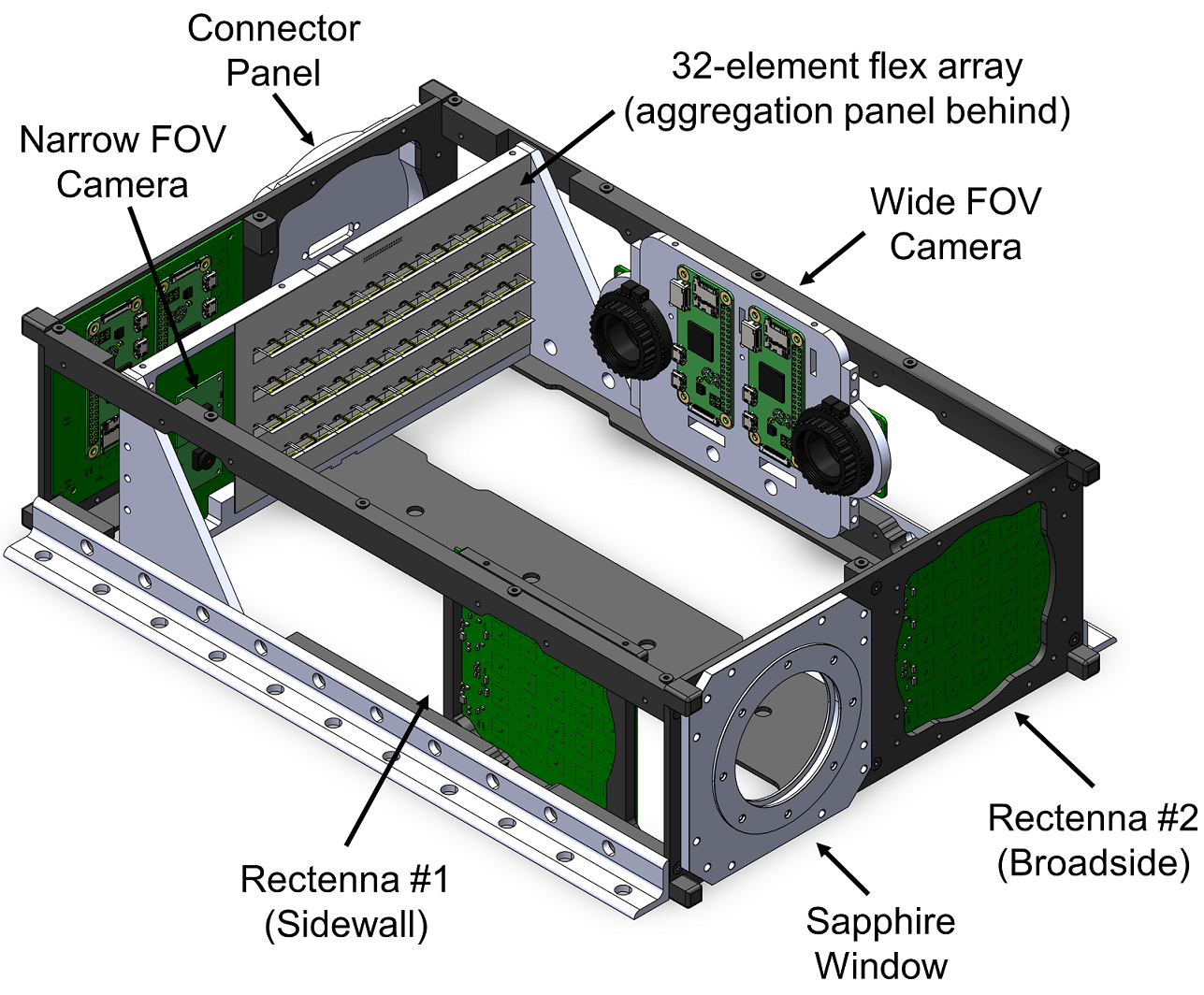}
   \caption{CAD rendering, labelled.}
   \label{fig:MAPLE_diagram} 
\end{subfigure}

\begin{subfigure}[b]{0.48\textwidth}
   \includegraphics[width=\linewidth]{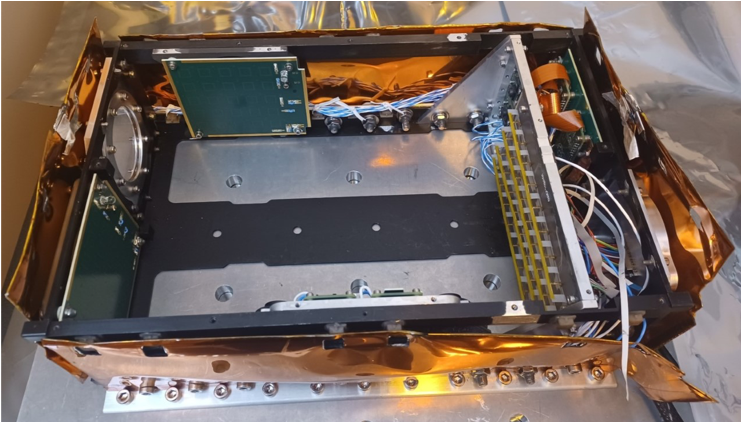}
   \caption{As assembled.}
   \label{fig:MAPLE_6u}
\end{subfigure}
\caption{Two depictions of MAPLE: a 6U CubeSat payload designed for testing and demonstration of in-space wireless power transfer.}
\end{figure}

\section{System Overview \& Design}
MAPLE, shown in Fig. \ref{fig:MAPLE_diagram}, is built around a 6U CubeSat chassis and serves to test and demonstrate wireless power transfer in space. MAPLE is mounted to the deck of Vigoride-5, a spacecraft which provided integration with the launch provider, communication, attitude control, and power supply infrastructure for MAPLE. Integration of MAPLE and Vigoride-5 is shown in Fig. \ref{fig:v5_integration}, taken before launch. The inside of the CubeSat, shown in Fig. \ref{fig:MAPLE_6u}, hosts a number of devices, including:
\begin{itemize}
    \item Two custom 16-channel silicon RFICs responsible for frequency generation, power amplification, and phase control
    \item An array of 32 flexible co-cured dipole antennas
    \item A flexible printed circuit board (PCB) on which the RFICs and antennas are mounted (``flex array" or FA)
    \item A rigid PCB with voltage regulation, on-board telemetry, and redundant microcontrollers that coordinate control of all of MAPLE's functionality (``aggregation panel" or AP)
    \item Two rectenna arrays with the option to load power to a resistor, bright LED, or a widebeam LED
    \item Two narrow field-of-view cameras
    \item Two wide field-of-view cameras
    \item An opening opposite the flexible PCB sealed with a sapphire window
\end{itemize}

\begin{figure}[t]
\centering
   \includegraphics[width=\linewidth]{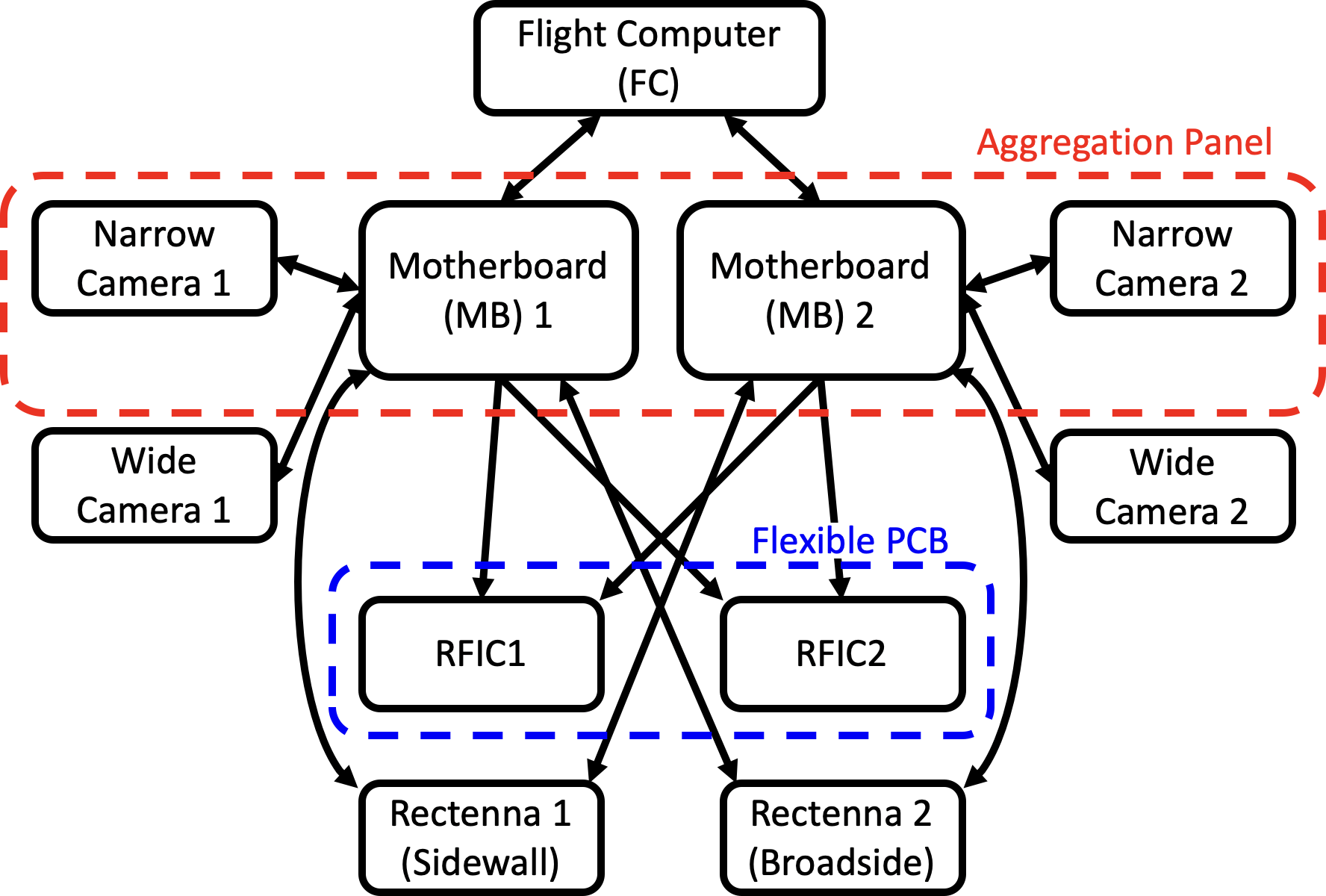}

   \caption{MAPLE block diagram, showing various components and component redundancy. Flight computer is not part of MAPLE, but has parallel communication lines and power lines to both motherboards on the AP.}
   \label{fig:MAPLE_components} 
\end{figure}

The payload is designed to maximize thermal and mechanical stability, given the requirement that it survive launch and months of thermal cycling in orbit. To this end, MAPLE was designed to sink heat from power sources to the flight deck while reflecting sunlight and hazardous ionizing radiation. Moreover, redundancy was built into every level of MAPLE. Every functionality was duplicated, allowing for the possibility of mission success despite possible component failure. This redundancy is evident and partially elucidated by Fig. \ref{fig:MAPLE_components}, which shows all of MAPLE's major components and the communication lines between them. In theory, most of the mission goals could be accomplished with the survival of only five out of 11 component blocks. Details on these designs are presented below.

\subsection{Hardware}
\subsubsection{Custom RFIC}
The core functionality of the WPT array is provided by the custom RFIC. The chip, designed in a 65nm CMOS process, provides the frequency synthesis, timing, beam forming and DC to RF conversion for 16 individual channels. Modern CMOS scaling enables this complex system to be integrated into a small, extremely lightweight $2.8\times2.8$mm die, a scale unattainable using discrete components. Two of these chips are used in MAPLE to drive the 32-element phased array.

\begin{figure*}[t]
\centering
   \includegraphics[width=\textwidth]{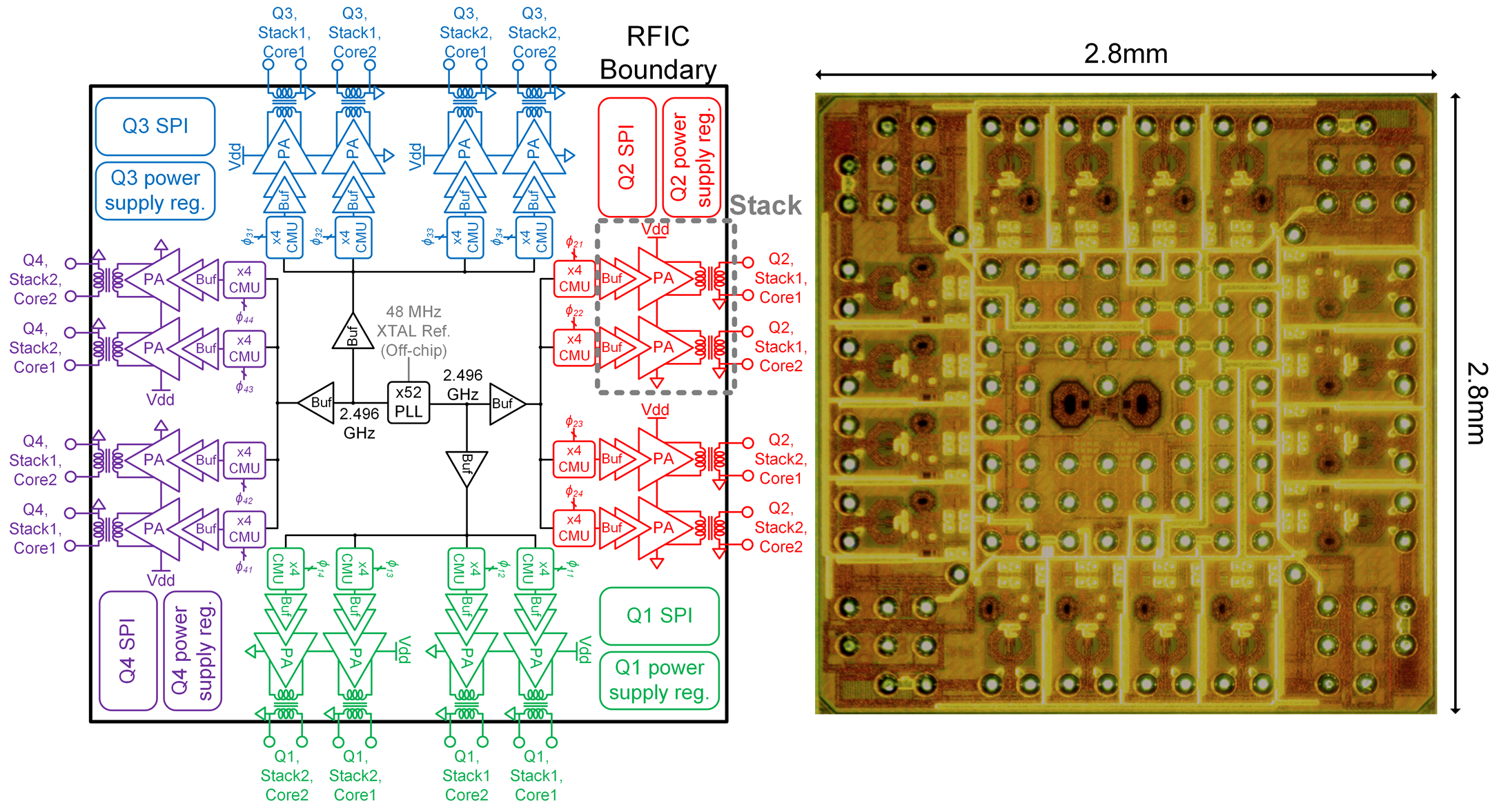}
   \caption{RFIC block diagram (left) and die photo (right).}
   \label{fig:RFIC_block} 
\end{figure*}

A block diagram and die photo of the chip is shown in Fig. \ref{fig:RFIC_block}. A 48MHz reference provided by an off-chip crystal oscillator is routed to both chips in MAPLE. On chip, it feeds into a centralized, x52 multiplication ratio phase locked loop (PLL) which synthesizes a 2.496GHz signal that is distributed to four quadrants (quads). Each quad consists of four cores. Each core contains a power amplifier chain (PA) driven by a fixed x4 multiplication ratio clock multiplier unit (CMU) that synthesizes the 9.984GHz local oscillator (LO). Phase shifting is accomplished by injecting a digitally programmable current into the loop filter of each core-level CMU \cite{hashemi2019flexible}. Adjacent PAs are powered in a series-stacked configuration, enabling higher supply voltages \cite{Abiri_XBandRFIC}. Each differential PA output is coupled to a 50$\Omega$ microstrip transmission line on the flex-array via an on-chip balun, which terminates at the pop-up dipole antenna. 

\subsubsection{Co-Cured Popup Antennas}
Implemented in this system are the flexible and lightweight co-cured popup antennas previously discussed in \cite{APSURSI_FIKES, mizrahi_popup,mizrahi_popup_TAP_2024}. These antennas, arranged in this implementation in four rows of eight elements each, are designed to be flexible enough that they can roll up with the board they are mounted on, light enough that they minimize launch costs, and durable enough to handle the harsh environment in space. These goals, often contradictory, are accomplished by the integration of the electronic and structural layers through ``co-curing," as presented in \cite{APSURSI_FIKES, mizrahi_popup,mizrahi_popup_TAP_2024}. The entire combination of electronic and composite layers are aligned, laminated, molded, and cured {\it together} - nontraditional in the context of electronics. This sidesteps the use of additional mounts or adhesives that can lead to mechanical weakness, mass, and/or electromagnetic complexity. It also allows for a simpler, more accurate, and more scalable 2D assembly.

The antennas were tested for electromagnetic and mechanical performance, especially as it relates to survival in harsh space environments. Results presented in \cite{mizrahi_popup, mizrahi_popup_TAP_2024} demonstrate the antennas possess all the typical advantages of a ground-backed dipole antenna (efficiency, bandwidth, simplicity, directivity, etc.) while also boasting extreme thermal and mechanical durability. The ground plane-backed dipole antennas with a vertical feed that doubles as a single-ended to differential balun were mounted to a single-ended transmission line in MAPLE's flexible PCB using 3M DP2216 epoxy.

\subsubsection{Flex Array} 

\begin{figure}[t]
\centering
   \includegraphics[width=85mm]{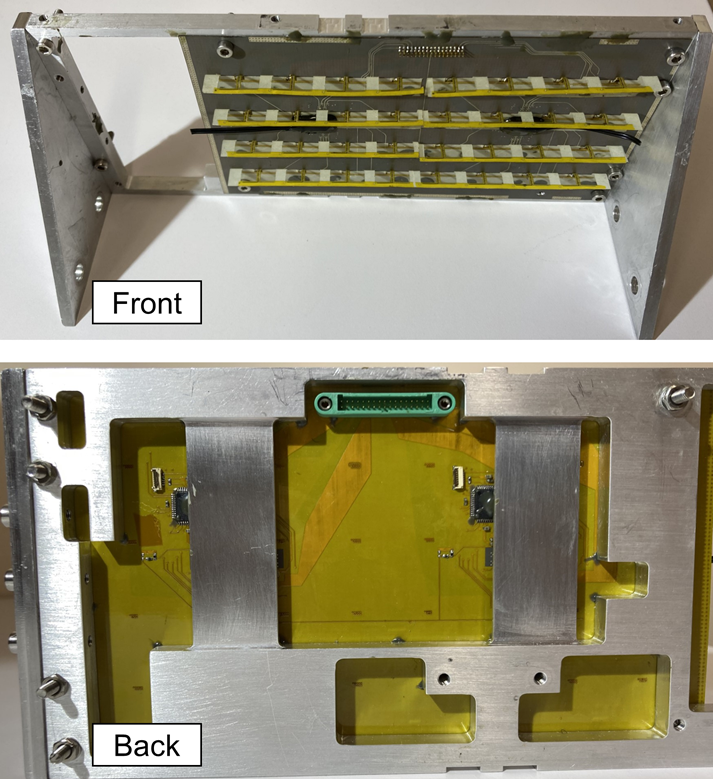}
   \caption{Flexible array with popup dipoles (engineering model depicted), fixed to an aluminum frame for heatsinking and mechanical integration.}
   \label{fig:flex_array_with_antennas} 
\end{figure}

The RFICs and popup antennas are mounted on a lightweight, flexible, four-layer polyimide PCB, which together are referred to as the ``flex array'' or FA. The FA measures roughly $160 \times 90$mm and, while physically a single board, effectively consists of two adjacent daughter boards, or DBs, each containing 16 antennas arranged in a $4 \times 4$ grid. Each daughter board is driven by a single RFIC, with digital control originating from an independent micro-controller mounted alongside. By radiating coherently, the two daughter boards form a phased array of 32 elements in an $8 \times 4$ configuration (Fig. \ref{fig:flex_array_with_antennas}). Considered together, the two sub-arrays and their RFICs increase the aperture size and, therefore, the focusing performance of the array; considered separately, they introduce an additional layer of redundancy.

The four metal layers provide the following functionality:
\begin{enumerate}
    \item Layer 1 provides component pads for the RFICs, microcontrollers, and associated passive circuitry.
    \item Layer 2 contains power distribution lines for 3.3V, 1.8V and 1.2V supplies that power the digital, power amplifier, and analog circuitry, respectively.
    \item Layer 3 consists of a low-mass ground plane for the microwave traces and antennas. The mass reduction relative to a standard ground plane is accomplished using a low-fill factor, rectangular mesh.
    \item Layer 4 routes 50$\Omega$ microstrip transmission lines to the antenna solder pads.
\end{enumerate}

\subsubsection{Aggregation Panel}
The Aggregation Panel PCB (``Aggregation Panel" or AP) provides the electronic interface between devices in MAPLE and the flight computer (FC) on Vigoride-5. It is a six-layer, conformally-coated, rigid polyimide board.

At the core of the AP are two redundant mother boards (MBs) (again, physically on the same board), wired in parallel, and each capable of communicating and controlling all subsystems, including the DBs and RFICs. Each MB consists of an Atmel SAMD21 microcontroller, and other accompanying devices. The AP also hosts all the necessary switches and sensors for system telemetry (current, voltage, temperature), connected to and controlled by the MBs. 

The AP interfaces with the two rectennas via two separate voltage sensing lines for the two MBs and control lines for switching between the rectenna LEDs and loads. The AP also hosts two narrow-FOV cameras and communicates with the wide-FOV cameras located off-board. Redundant crystal oscillators mounted on the AP provide the RFICs with a low-frequency reference; oscillators are powered and switched by either MB.

Finally, the AP hosts the voltage regulation devices which source power from the FC and output supplies to all MAPLE devices. All supply lines from the FC are 2x redundant; diode OR-ing of the FC supplies allow for operation with either supply in the event one fails.

\subsubsection{Rectennas}

\begin{figure}[t]
\centering
   \includegraphics[width=\linewidth]{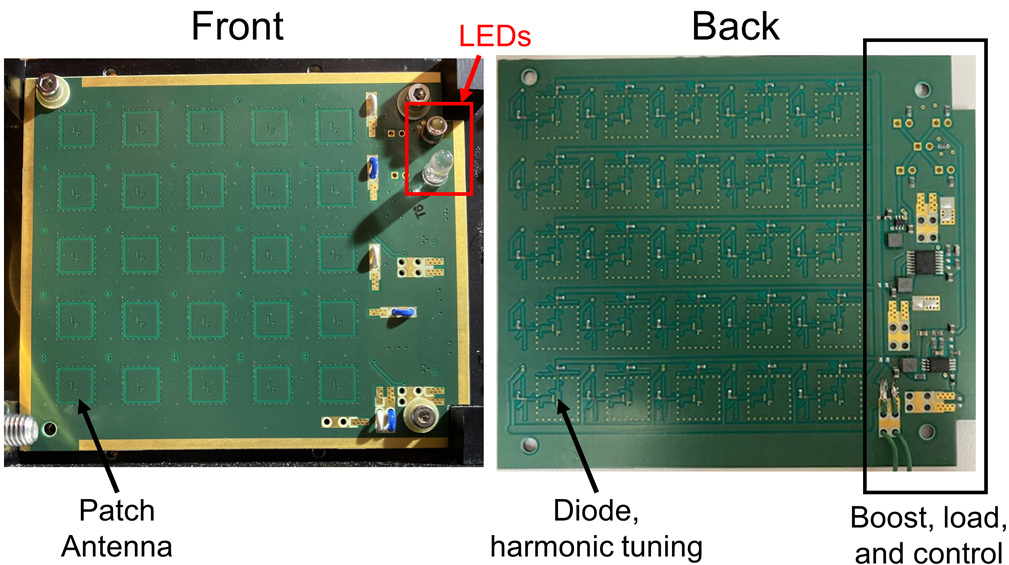}
   \caption{$5\times 5$ rectenna array of dual-polarized patch antenna receivers.}
   \label{fig:Rectenna_Picture} 
\end{figure}

MAPLE contains two rectennas that receive and convert the FA's wirelessly transmitted RF power to DC. Two boards provide redundancy, as well as spatial diversity to permit demonstration of the RFIC's ability to quickly steer power to different locations. One rectenna board is positioned broadside the array, while the other is offset and mounted on a sidewall. Rectennas are multi-substrate four-layer boards using Rogers 4003C for the antennas, Rogers 4350 for the rectifier circuits, and an FR4 pre-impregnated layer in the middle. 

Each rectenna comprises a $5\times 5$ grid of rectenna elements (combined in DC), two redundant maximum power point tracking boost circuits, two $22\Omega$ DC loads\footnote{Four individual, $22\Omega$, $\sfrac{1}{4}$W resistors are wired in a series-parallel configuration to increase the load power rating to $1$W and mitigate possible thermally-induced load variations.}, and two white LEDs each controlled by one of the two MBs (Fig. \ref{fig:Rectenna_Picture}.) Onboard MOSFETs allow a control signal from the AP to switch between a $22\Omega$ load and the boost circuit that powers the LED. The load serves both as a stable, power-optimal load for optimization as well as a safety precaution to prevent damage from open circuit rectenna operation. Each rectenna element consists of a dual polarized, probe-fed patch antenna that directly feeds two identical rectifier circuits, designed based on \cite{HAJIMIRI_DYNAMIC_FOCUSING}. For a nominal rectified power of 210mW, the rectenna array has a conversion efficiency of 40\% with respect to the power incident on its aperture. As shown in \cite{X_Band_High_Eff_Rectenna}, rectenna efficiencies at X-band can reach 70\% under optimum load and power settings.

\subsubsection{Cameras}
Four redundant cameras (two wide FOV and two narrow FOV) are present on MAPLE. Wide FOV cameras have $\approx 170^\circ$ FOV fisheye lenses attached to Raspberry Pi High Quality Cameras and are mounted on MAPLE's sidewall. Narrow FOV cameras (Raspberry Pi Camera Module 2) are mounted directly to the AP. Each camera is connected to a Raspberry Pi Zero W which handles camera operations and saves pictures to SD cards. UART serial lines provide the control and data transfer mechanism between the Raspberry Pis and AP.

\begin{figure*}[t]
\centering
   \includegraphics[width=\linewidth]{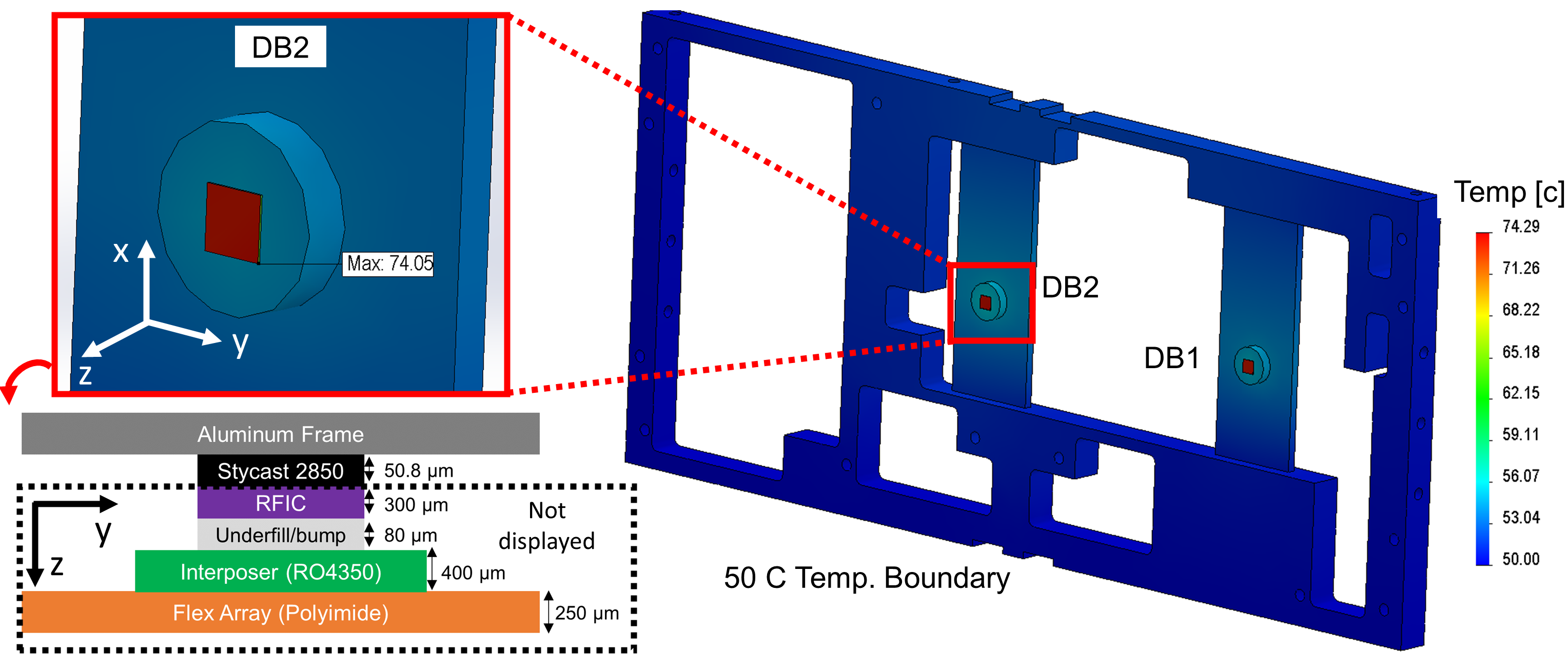}
   \caption{RFIC thermal simulation. Heat flows from the chip through the aluminum frame to the temperature boundary at the bottom face. Heatsink stackup is shown in the lower left (dashed box not displayed in other images.)}
   \label{fig:big_thermal} 
\end{figure*}

\subsection{Software}
\subsubsection{Focusing Algorithms}
MAPLE utilizes an implementation of the algorithm presented in \cite{HAJIMIRI_DYNAMIC_FOCUSING}. This algorithm is designed to bypass the trade-off between accuracy and runtime that typically plagues optimizations of extremely high-dimensional systems\footnote{For an array of 32-elements, each digitally controlled with a 9-bit phase code, there are $\left(2^{9}\right)^{32} = 512^{32} \approx 10^{87}$ unique phase configurations.}. It does this by sweeping over successively small ranges of phases with a large number of elements. Groups of elements swept together (called ``masks") help ensure changes exceed the noise floor of the system while also accelerating the optimization process by parallelizing tests. Orthogonal mask sets are chosen such that different combinations of elements are varied in different sweeps, thus increasing the search space explored in few iterations.

The focusing algorithm is predicated on a closed-loop feedback configuration of a ``generating unit" (GU) and ``receiving unit" (RU), as depicted in Fig. 5 of \cite{HAJIMIRI_DYNAMIC_FOCUSING}. In MAPLE, the FA (flexible PCB, antennas, and RFICs) constitutes the ``GU array." The array is controlled by a microcontroller on the MB which coordinates execution of the focusing algorithm by adjusting phases sent to the RFICs; the FA together with the MB constitute the GU. The MB also reads back voltage levels from the receiver array\footnote{Communication between the GU and the RU, though presented as using a wireless link in \cite{HAJIMIRI_DYNAMIC_FOCUSING}, was implemented in MAPLE using wired links.}, the RU, which is implemented as the rectenna discussed above.

MAPLE's optimization algorithm begins with an initialization of the RFIC settings and phases. Then, the MB begins adjusting select phases, detecting the resultant change at the receiver, and updating phases based on the change in received power. The algorithm continues like this for different combinations of elements until a specified number of iterations have been completed.

A functional system running the focusing algorithm would produce the type of curve, relating rectified power at the receiver to elapsed time, shown in Fig. \ref{fig:MAPLE_broad_opt}: power increases rapidly at the beginning and stabilizes at, presumably, a maximum.

Additionally, the ability to upload an array of digital codes from MB memory to the RFICs, herein called ``recall," was implemented. Recall allows MAPLE to measure the power rectified at the receiver using phase settings that were either saved in (nonvolatile) program memory (for example, to compare performance on Earth and in orbit) or in volatile memory (for example, to start an optimization with the phase codes output as a local maximum by a previous optimization during the same experiment.) The former enables MAPLE to make direct comparisons of system performance under identical phased array programming while the latter enables ``cascaded" optimizations that test whether a local maximum is actually a global maximum.

\subsection{Mechanical \& Thermal Design}

\begin{figure}[ht]
\centering
   \includegraphics[width=\linewidth]{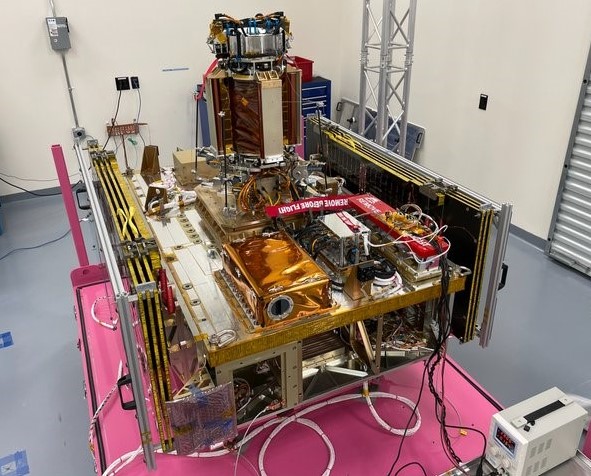}
   \caption{MAPLE integration on Vigoride-5 spacecraft from Momentus Inc. \cite{V5_integration}.}
   \label{fig:v5_integration} 
\end{figure}

\subsubsection{Mechanical Enclosure}
MAPLE's enclosure is a standard 6U aluminum CubeSat frame. The 6U frame contains all of MAPLE's electronics, as shown in Fig. \ref{fig:MAPLE_diagram}. Cabling for power and external communications are all provided by a single Micro-D connector mounted on an aluminum back-panel fixed to the CubeSat frame. The peripheral cameras and rectennas are fixed to aluminum rings which then mount to the CubeSat frame. A sapphire viewing port placed downrange of the FA is optically transparent for pictures and provides a seal to protect against atomic oxygen. The sapphire was chosen to be 4.4mm ($\lambda/2$ at 10GHz) thick to maximize its RF transmittance when focusing power outside of MAPLE. 

\subsubsection{Vigoride-5 Integration}

MAPLE was integrated on the Vigoride-5 spacecraft built by Momentus Inc. as shown in Fig. \ref{fig:v5_integration}. Vigoride-5 provides the power, communications, thermal sink, and attitude control for MAPLE as well as ALBA and DOLCE, two additional instruments in the SSPD-1 mission\cite{our_WiSEE_paper}. 

\subsubsection{Thermal Design}
All components in MAPLE are thermally coupled to the deck of the Vigoride-5 spacecraft. The deck serves as an isothermal boundary in the accompanying analysis. Most components are thermally coupled to the deck via connections from the AP's ground plane to the CubeSat frame. 

As the most concentrated sources of heat, the RFICs required special attention to ensure their temperatures did not rise above their effective operating point of 80$^\circ$C. The design and simulation of the RFIC's heat sink solution is shown in Fig. \ref{fig:big_thermal}. An aluminum frame that supports both the FA and AP includes two pedestals placed directly behind RFICs. The RFICs are bonded to these pedestals with Stycast 2850 thermal epoxy. The heat from the chips couples to Vigoride-5 predominantly through this aluminum frame. Thermal simulations with a 50$^\circ$C temperature boundary (worst-case) at the deck produces a max temperature of approximately 74$^\circ$C at the RFICs. 

In addition to the conductive heat sinking, MAPLE is enclosed by a 12-layer blanket of multi layer insulation (MLI) to prevent excessive radiative cooling and shield it from direct sun exposure. The outer layers of the MLI stackup are aluminized polyimide, while the inner layers are low emissivity aluminum blankets with perforations for air egress during the launch.

\subsection{Space Qualification Testing}
\subsubsection{Radiation Testing}
Radiation exposure is one of the primary, unique environmental challenges in space missions. Before launch, it was important to verify that all chosen components would survive radiation exposure for prolonged periods of time.

Commercial off-the-shelf (COTS) electronics were preferred due their low-cost and ease of acquisition. Compendiums of COTS parts subject to radiation testing at NASA's Jet Propulsion Laboratory (JPL) guided the selection process to ensure components had moderate resiliance to radiation events \cite{JPL_Rad_2015}\cite{JPL_Rad_2017}. For parts, such as the RFICs without space or radiation testing legacy, extensive proton radiation testing was performed. This testing ensured survivability under the mission's expected total ionizing dose. Parts were exposed to a 45 MeV proton beam for either one or three minutes at a time with a fluence of $10^{10}$ particles/$\text{cm}^2$. No physical or electronic damage was observed.

\subsubsection{Vibration Testing}
MAPLE was subjected to vibration testing as required by the launch provider, SpaceX, in accordance with their known launch acceleration spectrum. Vibration testing was performed a number of times over a two year period, both to individual components \cite{mizrahi_popup} and to the entire assembled payload. The following tests were performed:
\begin{itemize}
    \item Shock: A time-domain test meant to simulate sudden accelerations associated with dropping or banging a component.
    \item Sine Burst: A time-domain test at a single (low) frequency meant to test high displacement vibration.
    \item Sine Vibe: A frequency-domain sweep in accordance with the rocket's spectrum.
    \item Random Vibe: A time-domain test which generates a random vibration waveform in accordance with the rocket's spectrum.
\end{itemize}
Tests were performed in all three Cartesian axes; MAPLE passed all testing without damage and with minimal variations to its mechanical spectrum. Tests did not alter electronic performance.

\subsubsection{Temperature and Vacuum Testing}
Thermal vacuum chamber (TVAC) testing verified MAPLE could survive the harsh thermal environment of space. Thermal testing was performed on individual components \cite{mizrahi_popup, mizrahi_popup_TAP_2024} and the entire assembled payload, both in vacuum and at ambient pressure.

For TVAC testing, MAPLE was placed in a vacuum chamber pumped down to an internal pressure of  less than $10^{-6}$ torr. Temperature control in the chamber was programmed to simulate the moderate and extreme temperature limits expected in orbit:

\begin{tabular}{l l}
 --- Predicted (power-off) temperature limits & $[\ \ \ \ 5, 15]^\circ$C\\
 --- Protoflight temperature limits & $[-10, 30]^\circ$C\\
 --- Extreme temperature limits & $[-30, 60]^\circ$C
\end{tabular}

MAPLE spent several hours off, ``soaking" at the extreme minimum and extreme maximum limits. Then, several square wave cycles with a 95 minute period held MAPLE at the protoflight limits. Normal operation command sequences were tested during both the hot and cold phases. Several final soaks at the extreme limits followed, ending with a 100 minute soak at room temperature.

The payload survived testing, as verified by visual inspection, electrical testing, and TVAC testing results.

\section{In-Orbit: Experimental Results}
MAPLE was launched into orbit on a SpaceX Falcon 9 rocket on January 3rd, 2023 UTC. MAPLE's orbital height was approximately 527 km with an inclination of 97.5$^\circ$ \cite{space-track}. After initial qualification testing of Vigoride-5 completed, MAPLE's first experiment was conducted on March 3rd, 2023 at 11:51 PM UTC. MAPLE was deployed for a total of ten months; testing took place during the latter eight months.

\begin{figure}[t]
\centering
\begin{subfigure}[b]{0.48\textwidth}
   \includegraphics[width=\linewidth]{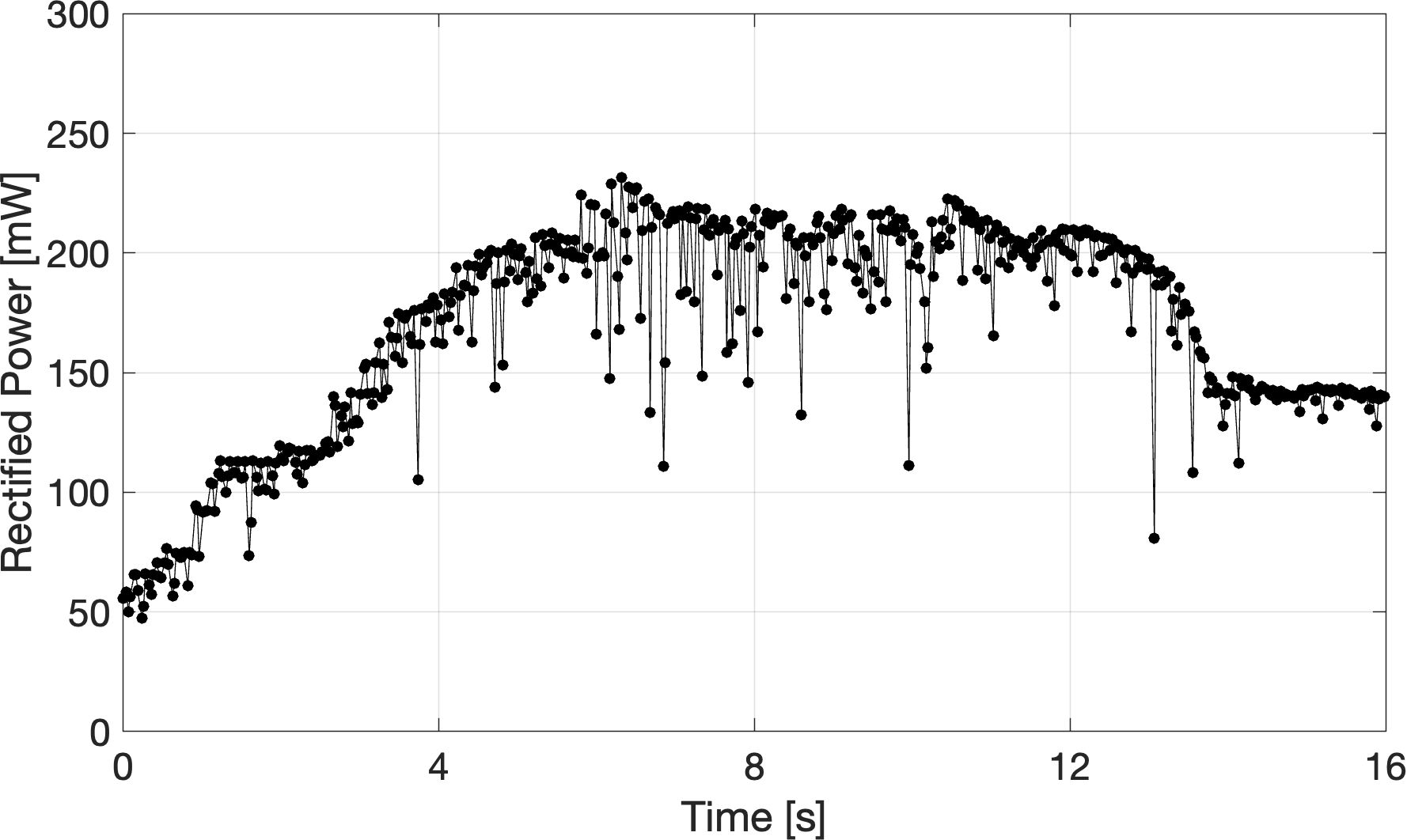}
   \caption{Rectenna \#1 (sidewall). Data from 2023/03/14, 03:19 UTC.}
   \label{fig:MAPLE_side_opt}
\end{subfigure}

\begin{subfigure}[b]{0.48\textwidth}
   \includegraphics[width=\linewidth]{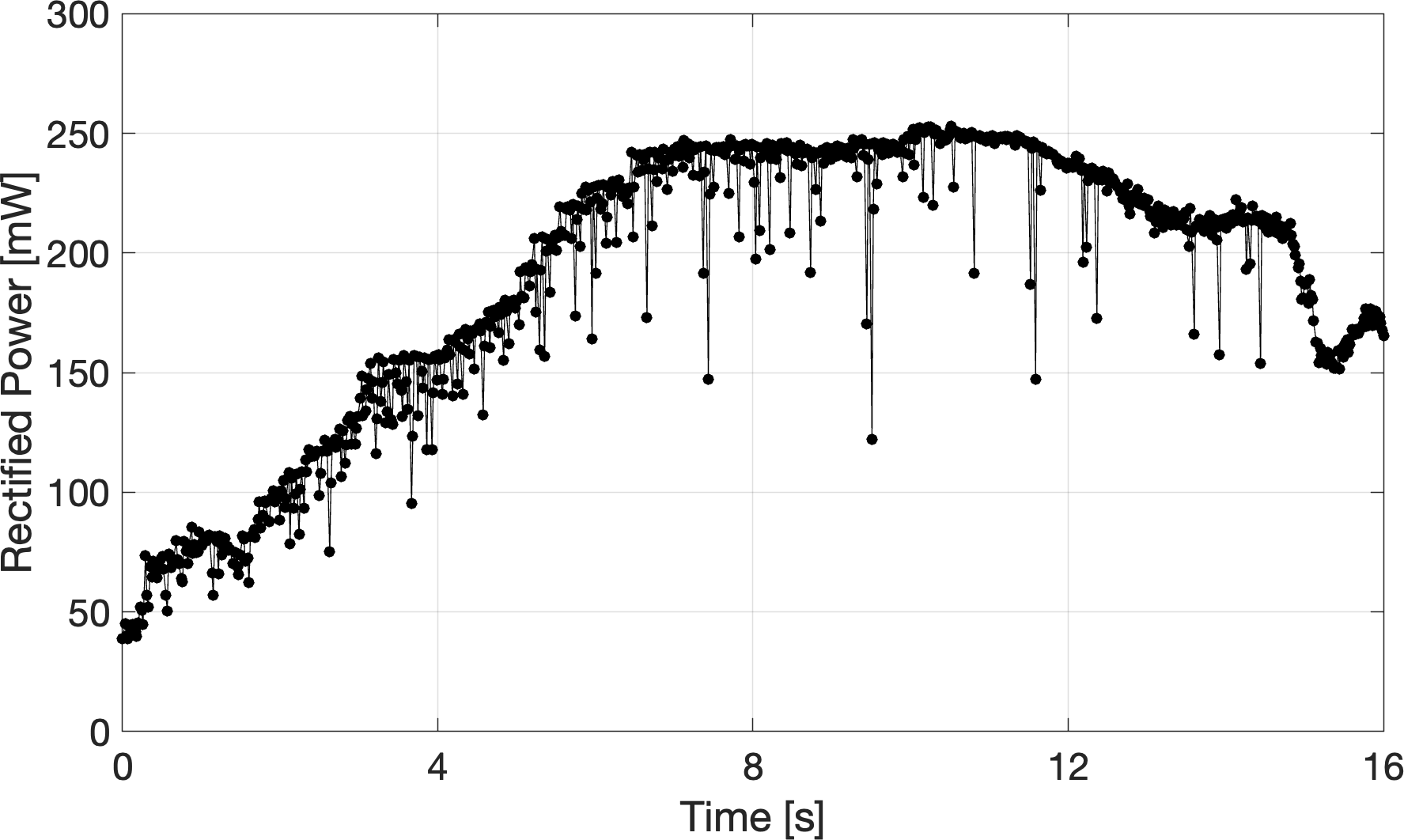}
   \caption{Rectenna \#2 (broadside). Data from 2023/03/03, 23:51 UTC.}
   \label{fig:MAPLE_broad_opt} 
\end{subfigure}

\caption{In-orbit focusing at rectenna \#1 and \#2 results.}
\end{figure}

\begin{figure}[t]
\centering
\begin{subfigure}[b]{0.48\textwidth}
   \includegraphics[width=\linewidth]{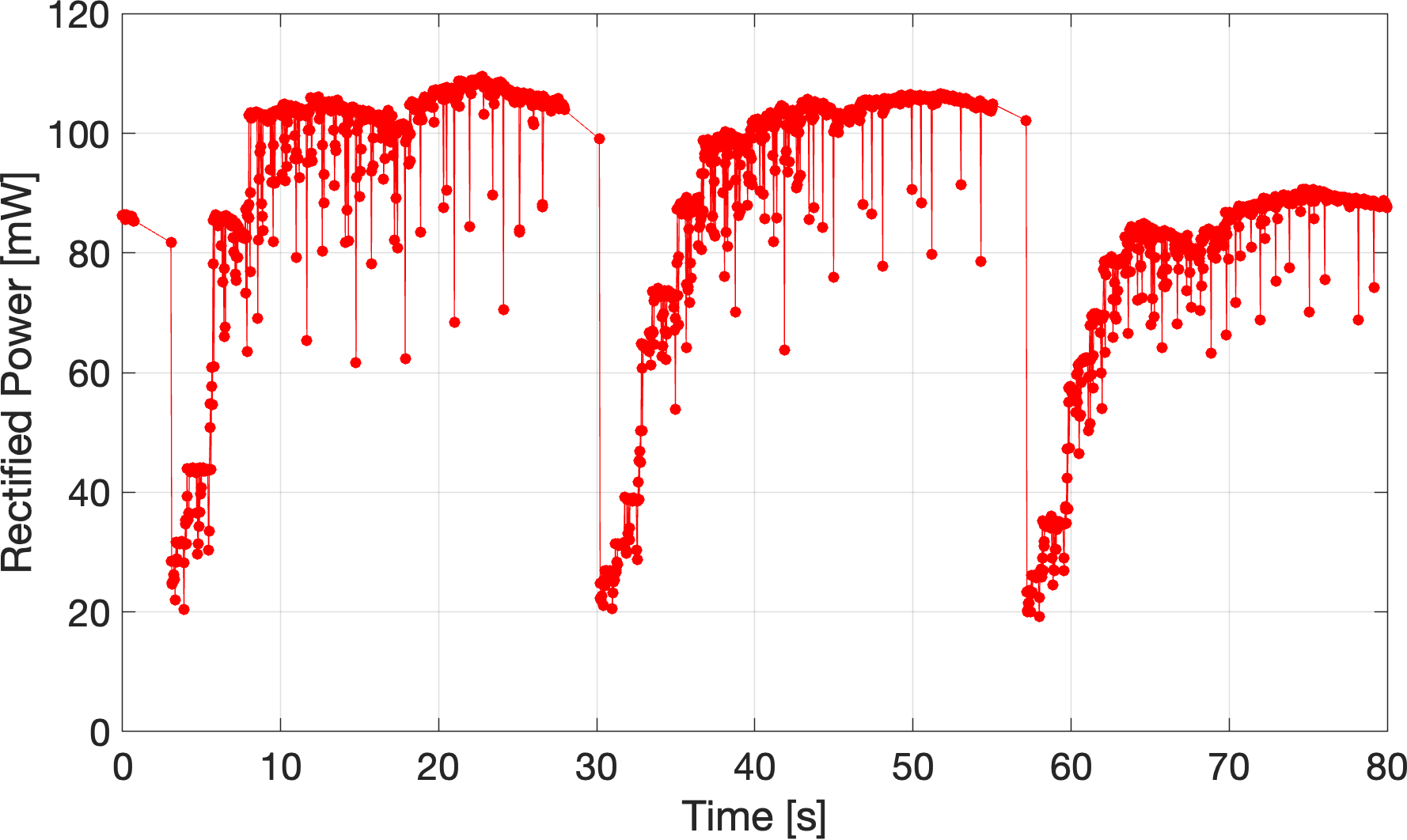}
   \caption{Rectenna \#2 focusing with only 16 elements controlled by DB1. Data from 2023/03/21, 23:16 UTC.}
   \label{fig:MAPLE_Just_DB1} 
\end{subfigure}

\begin{subfigure}[b]{0.48\textwidth}
   \includegraphics[width=\linewidth]{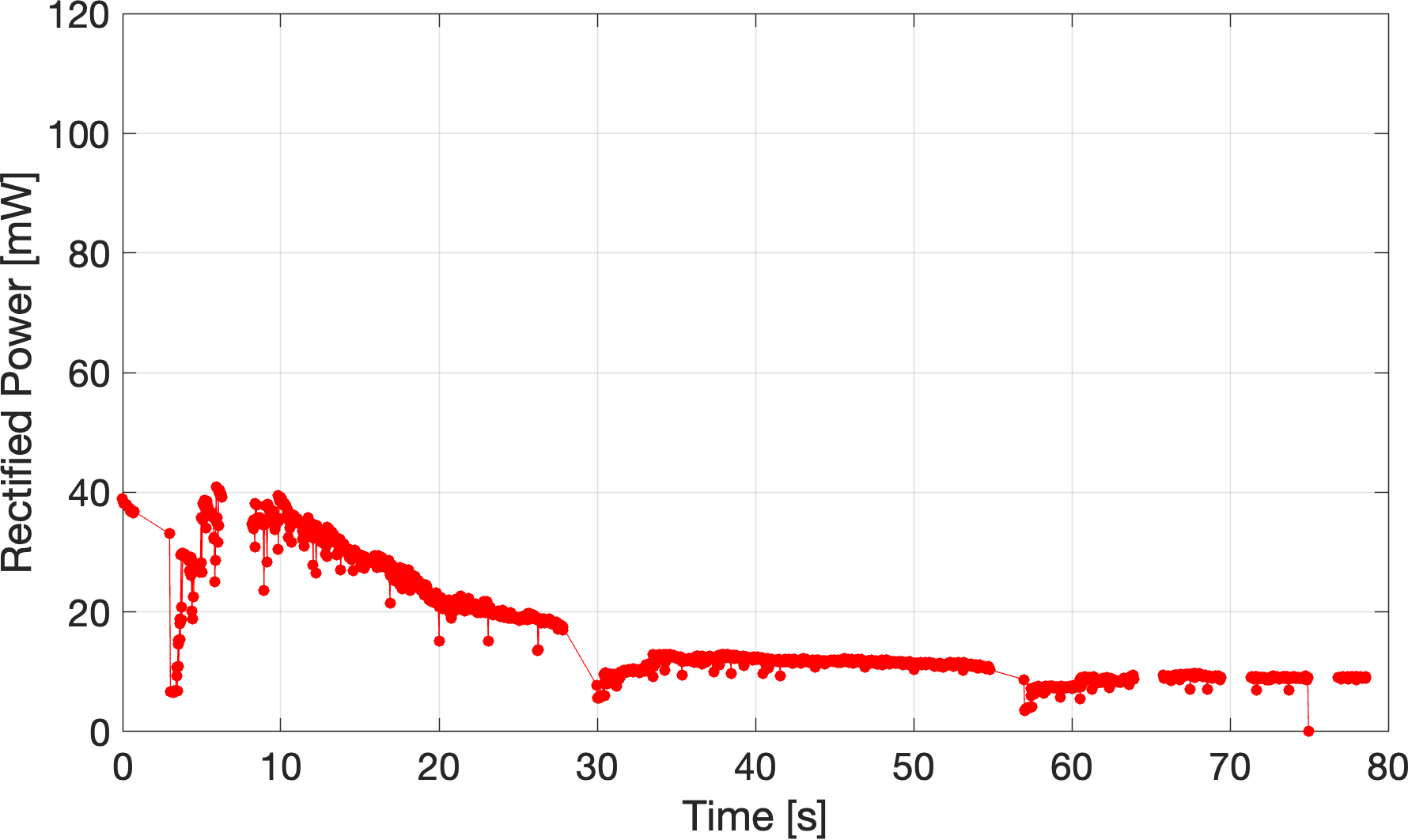}
   \caption{Rectenna \#2 focusing with only 16 elements controlled by DB2. Data from 2023/03/22, 00:47 UTC.}
   \label{fig:MAPLE_Just_DB2}
\end{subfigure}
\caption{In-orbit half-array focusing experiments.}
\end{figure}

\subsection{In-Orbit Wireless Power Transfer Experimental Results}
Effective execution of the focusing algorithm resulting in relatively high power received at the rectifier constitutes the single most important result from MAPLE. To successfully generate the desired optimization curve, all the software and hardware needs to work. Partial or complete failure of a component would affect several parameters of the measured optimization curves: the shape and/or the maximum asymptotic value.

\begin{figure*}[t]
\centering
   \includegraphics[width=\linewidth]{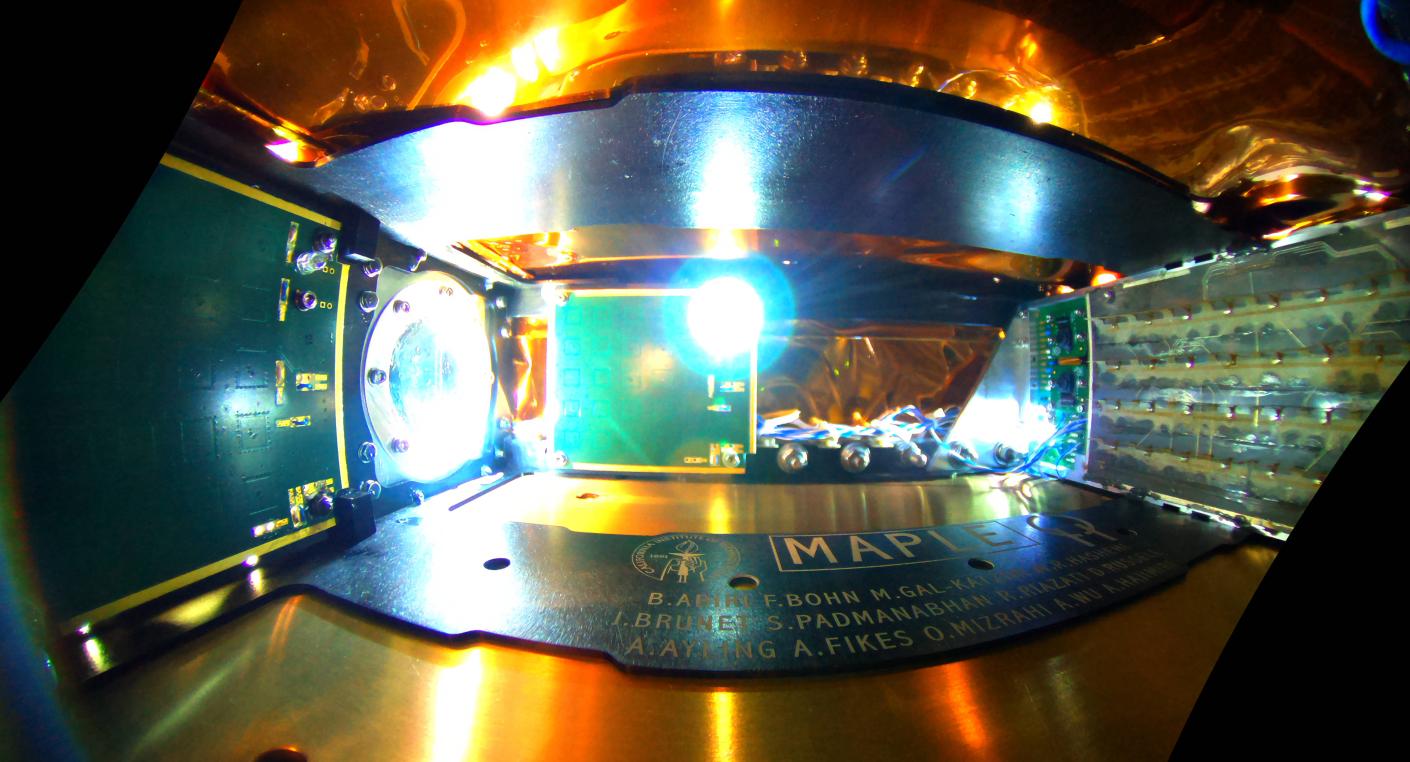}
   \caption{Interior of MAPLE during a power beaming experiment on March 15th, 2023 – 4:40 AM UTC. Lit LED (center) indicates power transmission to the (sidewall) rectenna.}
   \label{fig:LED_pic} 
\end{figure*}

The rectified power curves corresponding to optimizations performed in LEO are pictured in Fig. \ref{fig:MAPLE_broad_opt} and \ref{fig:MAPLE_side_opt}\footnote{The curves correspond to the optimizations that resulted in the highest rectified power to the respective rectennas during the duration of the mission.}. The flex array created coherent focal points on each of the rectennas. The peak rectified DC powers received by the the sidewall and broadside rectennas were 231mW and 251mW, respectively. Each of the 16 stacks, composed of two PA cores (Fig. \ref{fig:RFIC_block}), consumed about 230mA from a 1.73V supply. Once phases were optimized to maximize power rectified by the receiver, power was switched from a resistive load to a boost circuit which powers an LED. The intensity of the LED, shown in Fig. \ref{fig:LED_pic}, demonstrates the magnitude of power transferred.

Power degraded towards the end of the focusing in Fig. \ref{fig:MAPLE_broad_opt} and \ref{fig:MAPLE_side_opt}. This result is a deviation from ground testing. Separate in-orbit measurements focusing three times back-to-back using only 16 elements controlled by DB1 (Fig. \ref{fig:MAPLE_Just_DB1}) and DB2 (Fig. \ref{fig:MAPLE_Just_DB2}) were conducted. The results indicate degradation in the recovered power from DB2 that is exacerbated in subsequent optimizations. DB1, in comparison, output power consistent with ground tests and remained stable over time. Additionally, the 32 element power levels are restored once the system is restarted from an off state. Degradation in DB2 is the most likely cause of the drop in power seen in the full 32-element focusing experiments (Figs. \ref{fig:MAPLE_side_opt} and \ref{fig:MAPLE_broad_opt}.) More discussion of this phenomenon is presented in Section 5. 

\subsection{Individual Cores: Phase Shifting Performance}
MAPLE is equipped with a function which generates two-core interference patterns at the receiver for the characterization of individual element (core) performance.

\begin{figure*}[t]
\centering
    \includegraphics[width=\textwidth]{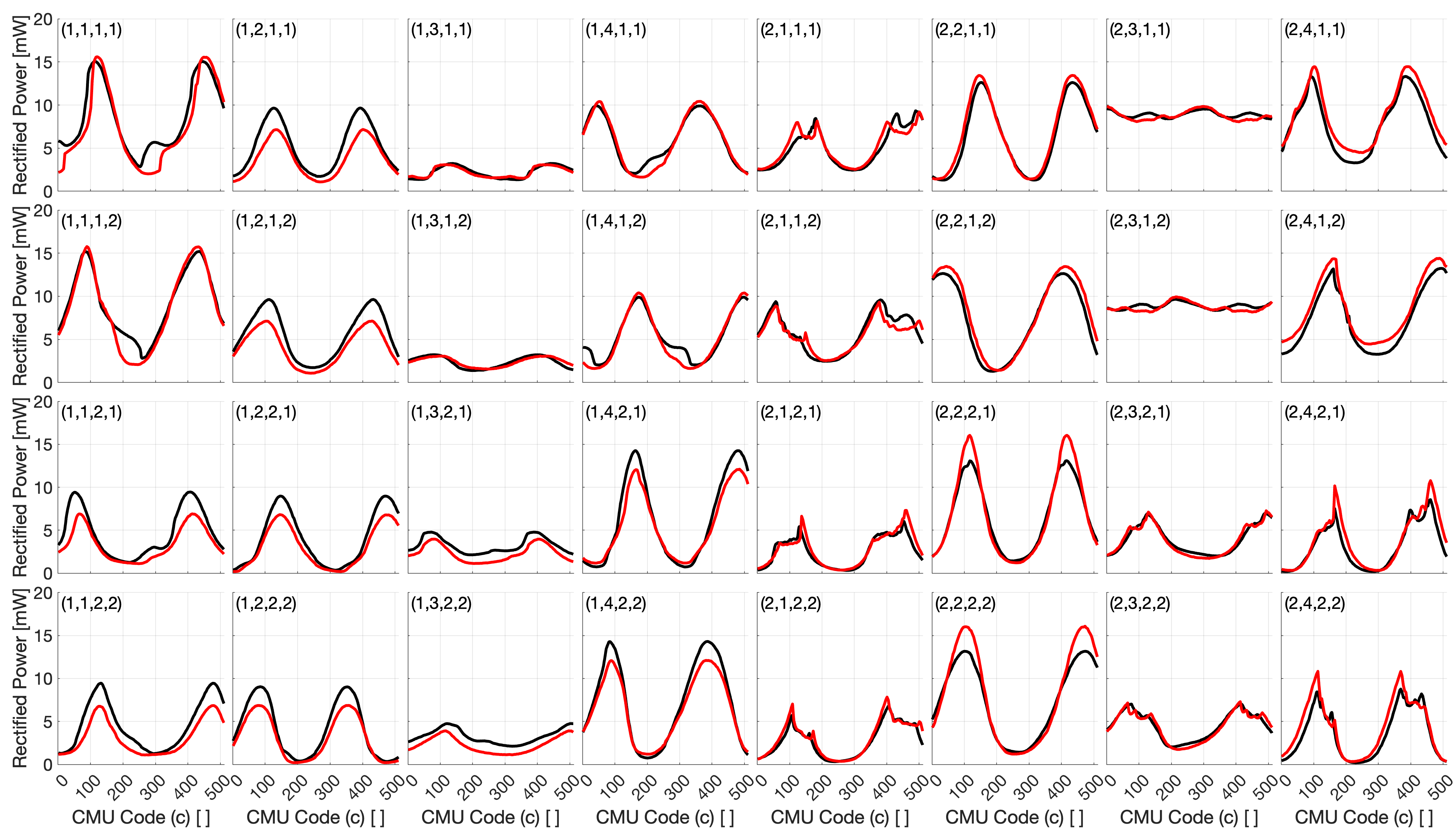}
    \caption{Raw rectified power from both on-Earth (black) and in-orbit (red) interference tests. On-Earth curves (black) represent the mean of six tests performed before launch; in-orbit curves (red) represent the first test performed in orbit. The label on the upper left indicates the ``coordinates" of the core: (RFIC, quad, stack, core). The y-axes have limits of [0 20] mW.}
    \label{fig:2core_both}
\end{figure*}

This testing of each of MAPLE's 32 elements proceeds as follows: addressing a single stack, one core is turned on with variable phase while the other is turned on with constant phase. The phase of the first core is swept digitally through some subset of the entire programmable range, set by (effectively) 9 bits of control\footnote{Phases are programmed using two bytes: a sign byte which is either all `1' or all `0' and a data byte. The sign byte only expresses an additional bit of information.}. At each phase iteration, voltage at the broadside receiver is collected and recorded back to the flight computer (FC) operating MAPLE. Then, the phase of the second core is swept with an identical subset of codes while the first core's phase is held constant. This is repeated for each of the 16 stacks across the two RFICs on MAPLE. An example of raw rectified power from one of these tests on Earth is presented in Fig. \ref{fig:2core_both}, in black. Data collected from tests in-orbit are overlaid in red.

Assuming a simplified, far-field propagation model and assuming a single phase center for the receiver, the electric field at the receiver during a phase sweep of element $m$ against constant phase on element $n$ can be modeled as follows:
\begin{align}
    \vec{E}_m(t,c) &= e^{j\omega t}\left(A_me^{j\phi_m(c)}P_me^{j\phi_m^P} + A_ne^{j\phi_n(0)}P_ne^{j\phi_n^P}\right)
\end{align}
where $\omega$ is the wave frequency, $\phi_m(c)$ is core $m$'s CMU phase function for code $c$, $n\in\{m-1,m+1\}$ is the index of the other core in the stack, $A_m$ is the amplitude of core $m$'s outgoing wave, and $P_me^{j\phi_m^P}$ is the propagation vector from core $m$ to the receiver (with amplitude $P_m$ and phase $\phi_m^P$). The propagation vector accounts for an ``infinite" sum of reflections emanating from the core which can vary owing (mostly) to perturbations on the MLI blanket and possibly other variable, internal physical features. This vector can vary, but is quasi-static for the measurement duration and likely changes negligibly in-orbit where there is no wind or (expected) physical change to the payload. The magnitude of the electric field can be expressed more simply as:
\begin{align}
    \left|\vec{E}_m(t,c)\right| &= \left|A_m'e^{j\phi_m'(c)} + A_n'\right|
\end{align}
where $A_m' = A_mP_m$ and $\phi_m'(c) = \phi_m(c) - \theta$ is the CMU phase function for core $m$, less a sum of constant phase terms, $\theta$.

When rectified, each phase sweep should produce a pseudo-sinusoidal interference pattern at the receiver. This interference pattern can be fit to the following model:
\begin{align}
    V_m(c) &= \sqrt{\frac{{A_m'}^2 + {A_n'}^2 + 2A_m'A_n'\text{cos}(\phi_m(c) + \theta)}{2}}
\end{align}
where $V_m(c)$ is the voltage read by the receiver when core $m$ is programmed to phase code $c$.

This model, $V_m(c)$, suggests a symmetry between the patterns generated by two cores in a stack: both patterns should be the same up to a constant phase shift and minor differences in the CMU phase function. This symmetry also prevents the determination of which core is $m$ and which is $n$\footnote{An alternative test has one core in an RFIC interfering with the other 15 cores programmed to a static phase shift. Asymmetry in the amplitudes virtually guarantees the correct identification of a core amplitude, but introduces other confounding factors: coupling between cores and RFIC heat generation associated with full 16-core operation which can alter or degrade core performance. This ``1 vs. 15" test is {\it not} used for results presented in this paper.}. By convention, we assign the larger of $\{A_m', A_n'\}$ to be $A_1'$ and the smaller to be $A_2'$. We cannot distinguish between the two amplitudes in a stack enough to assign them to their respective cores.

This model allows for the prediction of $A_m'$, $A_n'$, and extraction of the CMU function. Both the amplitudes and the CMU function are metrics indicative of RFIC performance and allow comparison of on-Earth and in-orbit performance.

\begin{figure*}[t]
\centering
    \includegraphics[width=\textwidth]{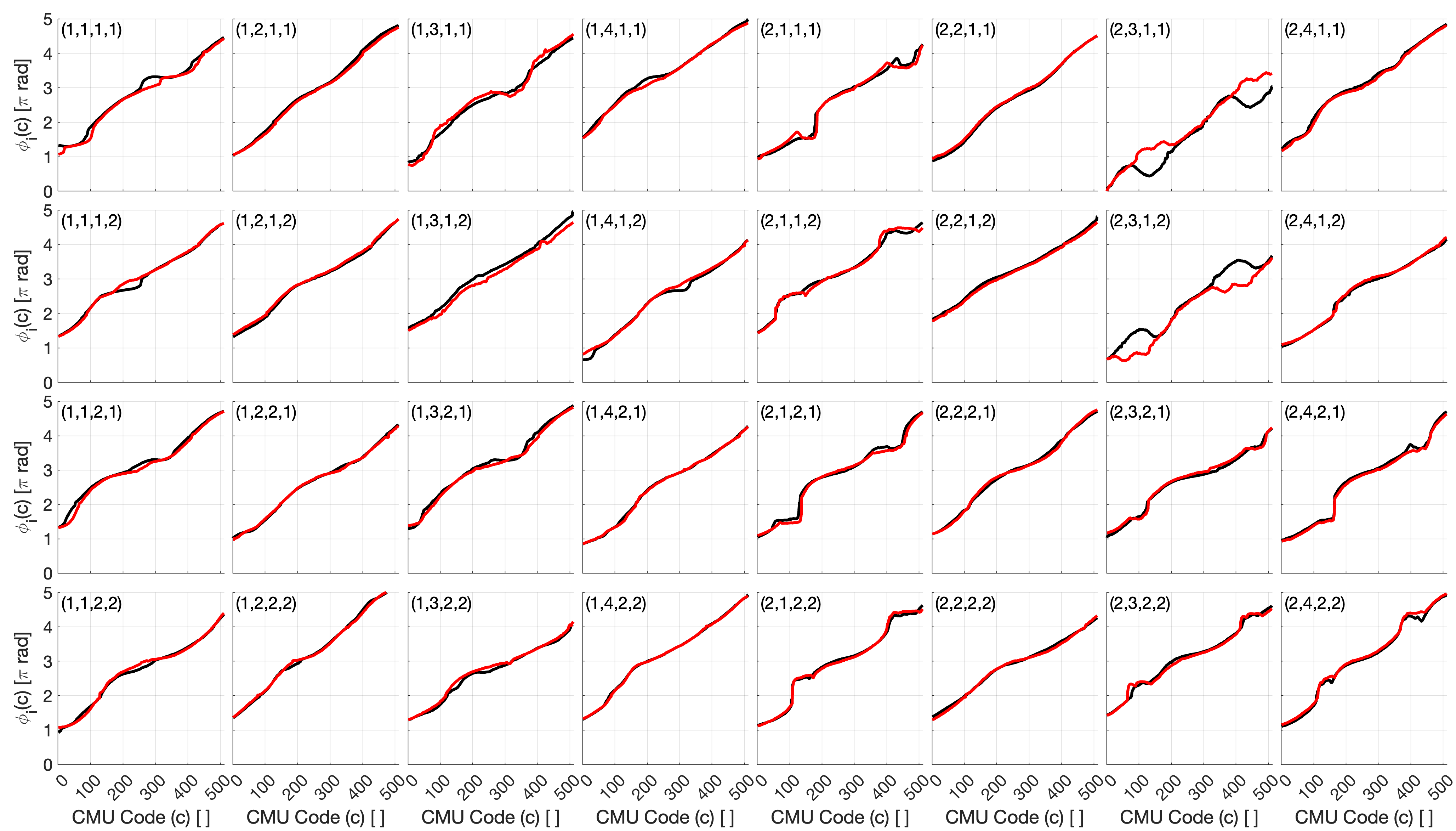}
    \caption{In black: unwrapped CMU phase functions, averaged over $N=6$ interference tests performed during (on-Earth) integration prior to launch, for each core. In red: unwrapped CMU phase functions from the first in-orbit test. The label on the upper left indicates the ``coordinates" of the core: (RFIC, quad, stack, core).}
    \label{fig:2core_CMU_unwrapped}
\end{figure*}

Six on-Earth tests performed after integrating MAPLE onto the flight deck of Vigoride-5 but before launch, herein called ``integration tests," are the best indications of on-Earth performance for the purpose of comparison to in-orbit data. Integration tests were performed as close to launch as possible to provide the best possible baseline for in-orbit performance. CMU phase functions from each of these tests were individually extracted by inverting $V_m(c)$ and unwrapped; the ensemble average of these curves over the six integration tests is shown in Fig. \ref{fig:2core_CMU_unwrapped}, in black. The CMU phase functions are all roughly linear and are capable, on average, of producing $\approx 3.37\pi$ rad of phase rotation over $512$ unique phase codes. Averaged over tests, cores, and phase codes, the absolute deviation from the mean CMU phase function for integration tests is small at $3.90^\circ$, demonstrating the CMU's ability to consistently produce the same output phase for a given input code.

\begin{figure}[t]
\centering
    \includegraphics[width=0.48\textwidth]{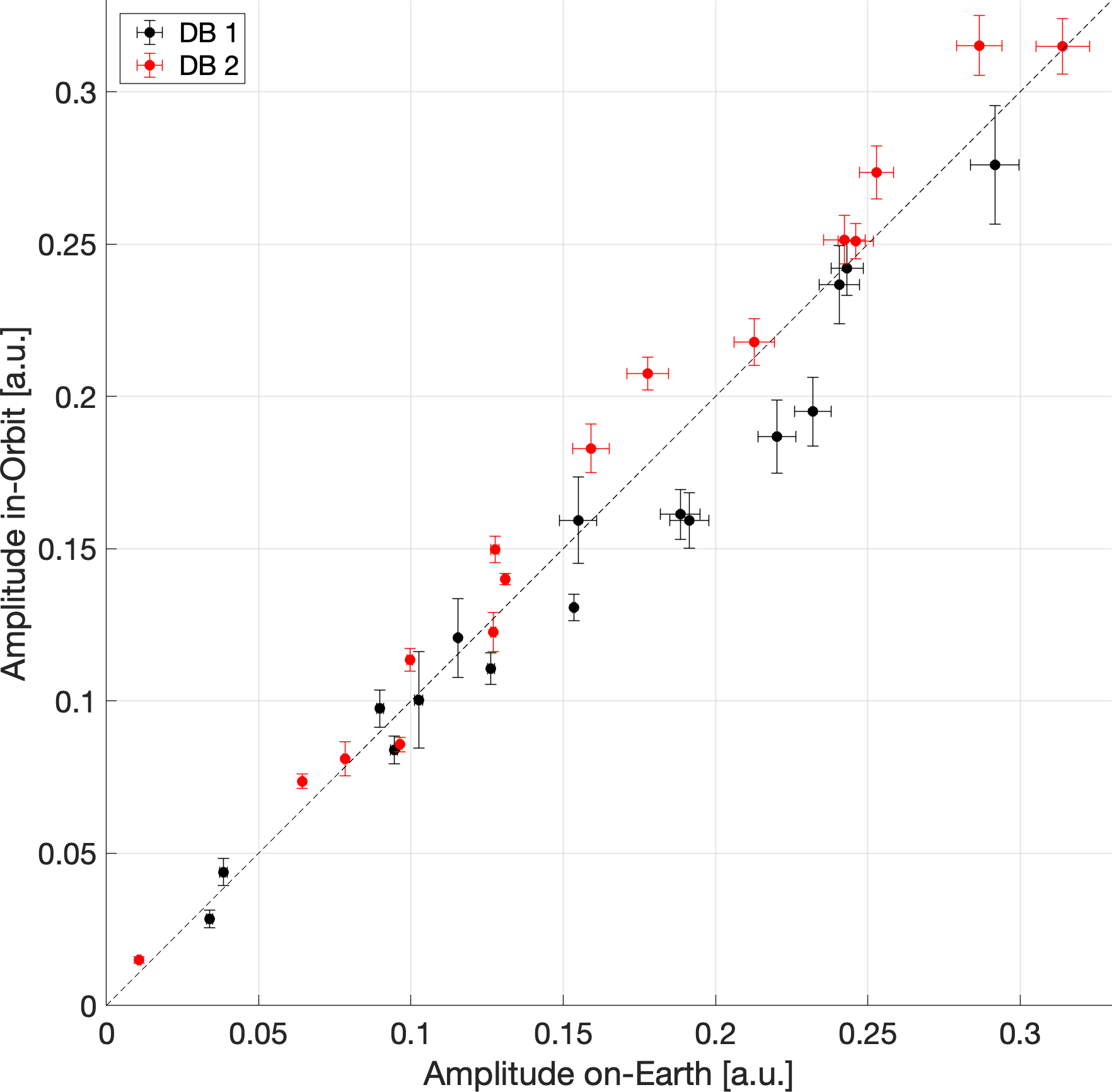}
    \caption{Comparing on-Earth to in-orbit PA amplitudes for 32 cores across two RFICs. Error bars indicate spread (standard deviation) among tests performed on-Earth (x-axis) and in-orbit (y-axis).}
    \label{fig:amplitudes_compare}
\end{figure}

PA amplitudes, as measured by $A_m'$ and $A_n'$, were consistent between tests, with standard deviation between tests, averaged over all cores, of 2.92\% of the mean for $A_1'$ and 1.35\% for $A_2'$. PA amplitudes varied between cores, with a standard deviation between cores of 19.77\% for $A_1'$ and 41.77\% for $A_2'$. Given that each antenna is positioned differently, each will interact with MAPLE's cavity differently. Thus, PA amplitude variance is possibly owing to variance between scattering vectors. A comparison between on-Earth and in-orbit amplitudes is given in Fig. \ref{fig:amplitudes_compare}.

Three in-orbit tests performed over two days, two months after launch, can be compared to integration tests and among themselves to understand how the CMUs perform differently in orbit and over time. The ensemble standard deviation of CMU phase functions among in-orbit tests, averaged over cores and phase codes, was $5.87^\circ$, greater than that for the integration tests. This is indicative of decreased CMU phase-shifting performance stability in space, possibly caused, as discussed below, by increased RFIC temperatures owing to the vacuum in space. Compared to the mean CMU curve from integration testing, in-orbit CMU curves have mean absolute deviation of $20.43^\circ$ but median absolute deviation of $9.72^\circ$. Thus, CMU phase-shifting stability is slightly worse in space, and the exact CMU curve changes as a result of the change in environment.

PA amplitudes in-orbit have a standard deviation among tests, averaged over cores, of 4.53\% for $A_1'$ and 6.31\% for $A_2'$. This is approximately double the variation in PA amplitude for on-Earth testing, possibly owing to varying thermal conditions. Average signed deviation, as a percent of on-Earth amplitudes, is -0.61\% for $A_1'$ and 2.49\% for $A_2'$, implying that no systematic shift (increase or decrease) in PA output power occurred as a result of in-space operation. Average absolute deviation, as a percent of on-Earth amplitudes, is 8.74\% for $A_1'$ and 12.73\% for $A_2'$, implying in-space operation resulted in variation in PA output power, in this case.

\subsection{Long-Term Performance Degradation}

\begin{figure*}[t]
\centering
    \includegraphics[width=\textwidth]{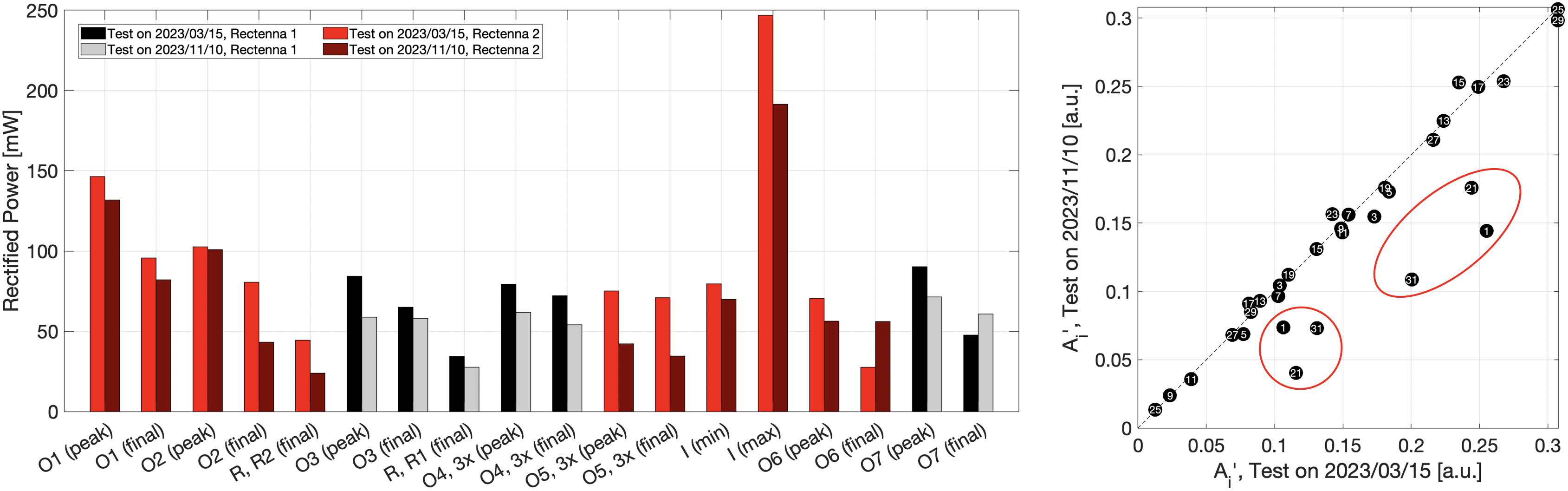}
    \caption{Tracking long-term performance degradation by comparing two benchmark experiment runs eight months apart. Left: Power rectified during different phases of the benchmark experiment. ``O" = optimization, ``R" = recall, ``I" = interference. ``O1 (peak)" denotes the peak power during the first optimization. Right: A comparison of individual core amplitudes ($A_i$) at the beginning (x-axis) and end (y-axis) of the ten month deployment. Each data point corresponds to a core, labeled using the core number ($\in [1,32]$) for the first core in the stack. Degraded cores circled in red.}
    \label{fig:before_after}
\end{figure*}

MAPLE ran a benchmark experiment sequence in the first and last days of its ten month deployment period. These two runs were performed under nearly identical conditions with a nearly identical sequence in an effort to directly quantify long-term system degradation. The sequence consists of a number of single optimizations, successive optimizations (one optimization starting where the previous stabilized), recalls, chip-to-chip interference cycles, and a characterization of individual core performance. Fig. \ref{fig:before_after} compares the rectified power during these tests. Considering a mean taken over these 18 benchmark values, power declined by 14.7\%.

However, when individual cores were analyzed, we observe degradation in performance in only three stacks (six cores), circled on the left of Fig. \ref{fig:before_after}. The other 26 cores demonstrated a very high correlation between amplitude in the early and late tests, appearing virtually unaffected by time spent in space. CMU phase functions (not shown) were virtually unchanged over the deployment period.

Summing the individual amplitudes and assuming ideal coherence, degradation in these three stacks contributes to a total degradation in {\it potential} peak power of 18.25\%. The similarity between measured decline in peak power and potential peak power based on individual core amplitudes (14.7\% vs. 18.25\%, respectively) suggests that power decrease is not owing to a {\it broad, chip-level} degradation in performance, but rather the effect of a number of cores that have individually degraded. Moreover, the system's ability to reach 191mW (I (max), 2023/11/10), close to the original peak of 247mW (I (max), 2023/03/15) prior to long-term degradation, suggests that fundamental system performance is largely maintained over the experiment lifetime.

\section{Beaming to Earth: Experimental Results}
As a demonstration of both MAPLE’s functionality and ability to guide its transmitted energy in the desired direction, a proof-of-concept experiment was performed wherein MAPLE beamed RF power to a ground station at Caltech in Pasadena, CA, USA where it was detected\footnote{``Power transfer" is defined as a net positive recovery. In contrast, ``detection" is defined as a process where the recovered power is less than the external power needed for the receiver.}.


In these experiments, MAPLE was scheduled to turn on, steer power out the sapphire window opposite the array, and maintain its state for several minutes while passing over Pasadena. The ground station used for detection during the experiment was assembled on the roof of Caltech’s Moore Laboratory. Discussed below are the ground station design details and results from the beaming experiment.


\subsection{Ground Station Design}
The ground station consists of two critical elements: an RF receiver chain and a tracking apparatus.

\subsubsection{RF Receiver}
MAPLE transmits a 9.984GHz continuous wave, sinusoidal signal with an EIRP of approximately between 26.5-28.5dBm (Fig. \ref{fig:antennaPat}). With a projected free space path loss of 167dB\footnote{167dB free space path loss is based on a payload directly overhead, at an orbital altitude of approximately 527km.}, the predicted incoming signal power, assuming ideal pointing coordination, is approximately -138.5dBm to -140.5dBm, implying a requirement for a high-gain receiver. A 33.5dBi gain, 2.8$^\circ$ half-power beamwidth, dual-polarized dish antenna served as the receiving aperture, providing high-gain while being small enough (67cm diameter) to fit on commercial telescope tracking mounts.

\begin{figure}[t]
   \includegraphics[width=0.48\textwidth]{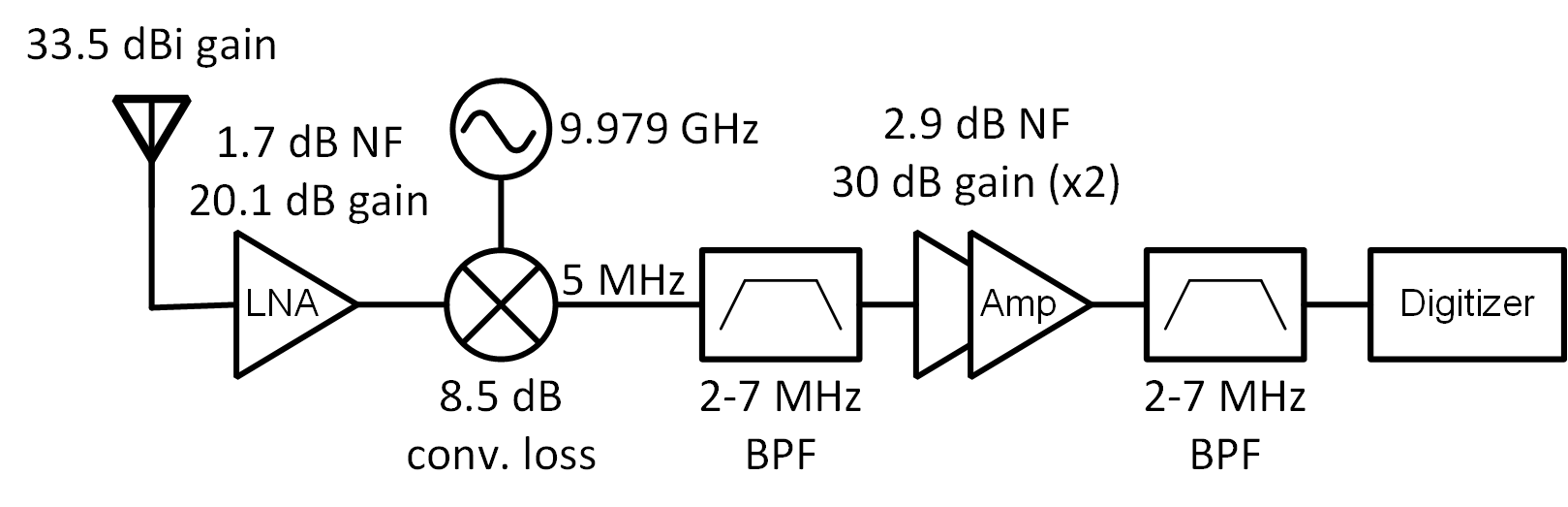}
   \caption{Block diagram of RF receiver chain for ground station. There is an additional 3 dB of loss due to cabling from the LNA to the mixer.}
   \label{fig:blockdiagram} 
\end{figure}

The super-heterodyne receiver, shown in Fig. \ref{fig:blockdiagram}, down-converted MAPLE's 9.984GHz signal to a 5 MHz IF signal which was then filtered, amplified, and digitized. A low-noise amplifier (LNA) was attached directly to the output waveguide transition on the antenna, before the mixer, to amplify the signal and reduce the noise contribution of subsequent stages. A 2-7MHz bandpass filter removed potential interference signals from the IF. The receiver has a gain of +68.5dB with a total noise figure of 5dB. Two identical receivers were used for each polarization of the dish antenna.

The IF signal from the receiver was digitized at 20Msps by a Keysight M3100A PXIe Digitizer module for additional processing and analysis. The received signal's SNR was maximized using a matched filter which, in this case, was a fast fourier transform (FFT) due to the near-sinusoidal nature of the signal. A processing interval of 26ms allowed for a sufficiently high SNR without introducing frequency smearing between FFT bins owing to a continuous Doppler shift. 16 IF spectra were produced every second throughout the pass.

\subsubsection{Tracking}
Tracking involved coordination between three individual entities: the timing of MAPLE's software operations, Vigoride-5's attitude, and on-Earth ground station tracking.

MAPLE operations were programmed in advance, timed to begin in coordination with the payload crossing the horizon relative to Pasadena, CA, USA. Timing accuracy was confirmed by other experiments and proved to be more than sufficient.

\begin{figure}[t]
    \centering
    \includegraphics[width=\linewidth]{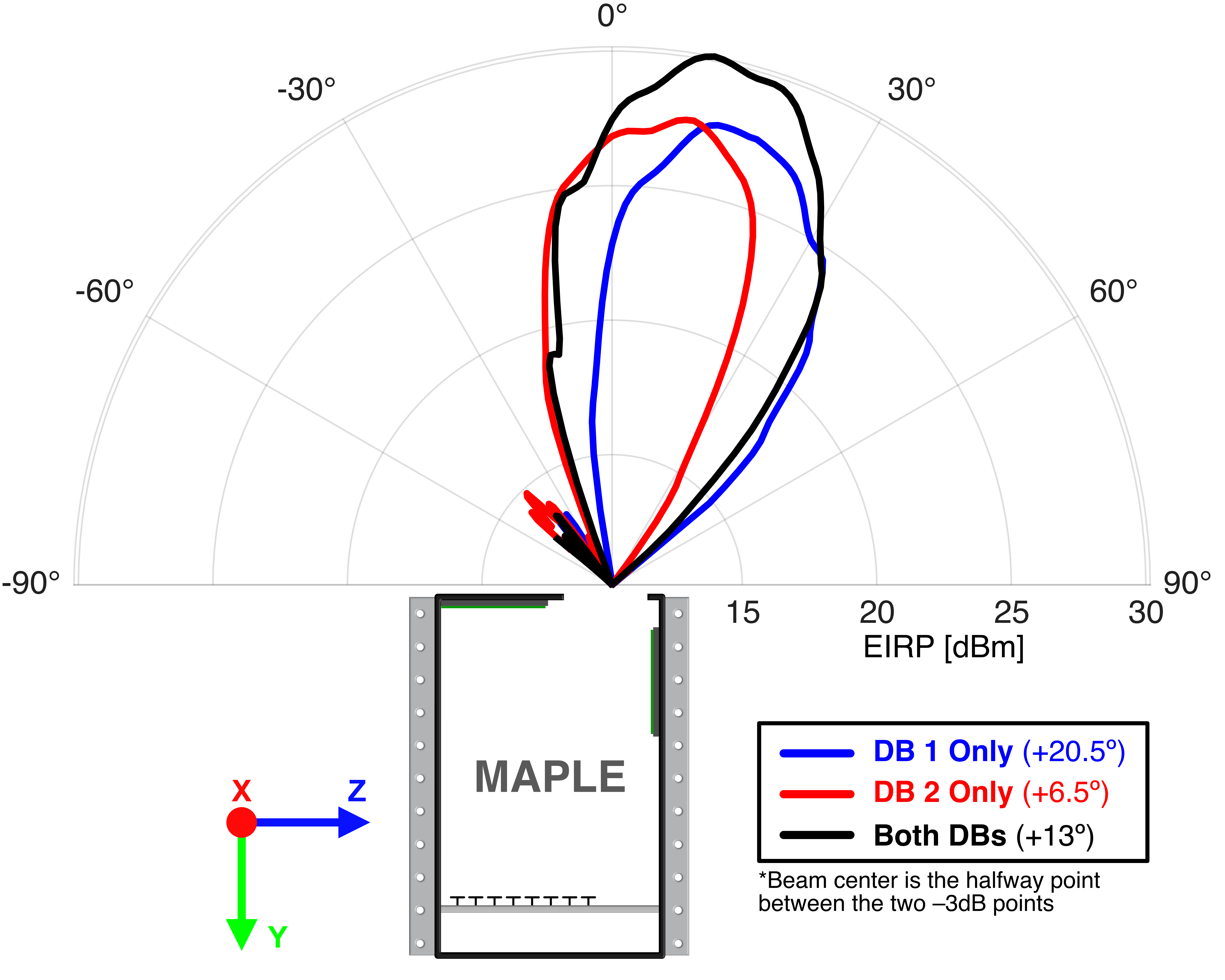}
    \caption{MAPLE antenna pattern through the sapphire window for radiation from DB1, DB2, and both DBs.}
    \label{fig:antennaPat}
\end{figure}

The 16-element subarrays driven by DB1 and DB2 are placed side by side in MAPLE; each combination of DB1, DB2, and DB1+DB2 corresponds to a slightly different main beam direction. These beams, characterized under nearly identical circumstances in an absorbing chamber on Earth, are shown in Fig. \ref{fig:antennaPat}. The main beam is not exactly broadside, owing to the relative skew of the viewing window and the selection of active RFICs. In all of the BTE experiments, only DB1 was powered on while DB2 remained off\footnote{Using only a single RFIC trades potentially higher power for higher stability owing to possible chip-to-chip interference. DB1 proved more stable overall, as shown in figures \ref{fig:MAPLE_Just_DB1} and \ref{fig:MAPLE_Just_DB2}.}. Thus, Vigoride-5 was programmed to accommodate a pointing correction of +20$^\circ$ relative to broadside so that MAPLE's main beam would be directed at the ground station.

To maximize received power from MAPLE on Earth, the ground station antenna continuously pointed at (``tracked") the payload during the pass. Because of the high-gain dish antenna's narrow beamwidth, pointing accuracy was critical to avoid a large power penalty for misalignment. To this end, the antenna was mounted on a modified telescope mount (the iOptron AZ Mount Pro) with a sufficiently high peak slew rate of $10^\circ/\text{sec}$. 

\begin{figure}[t]
   \includegraphics[width=0.5\textwidth]{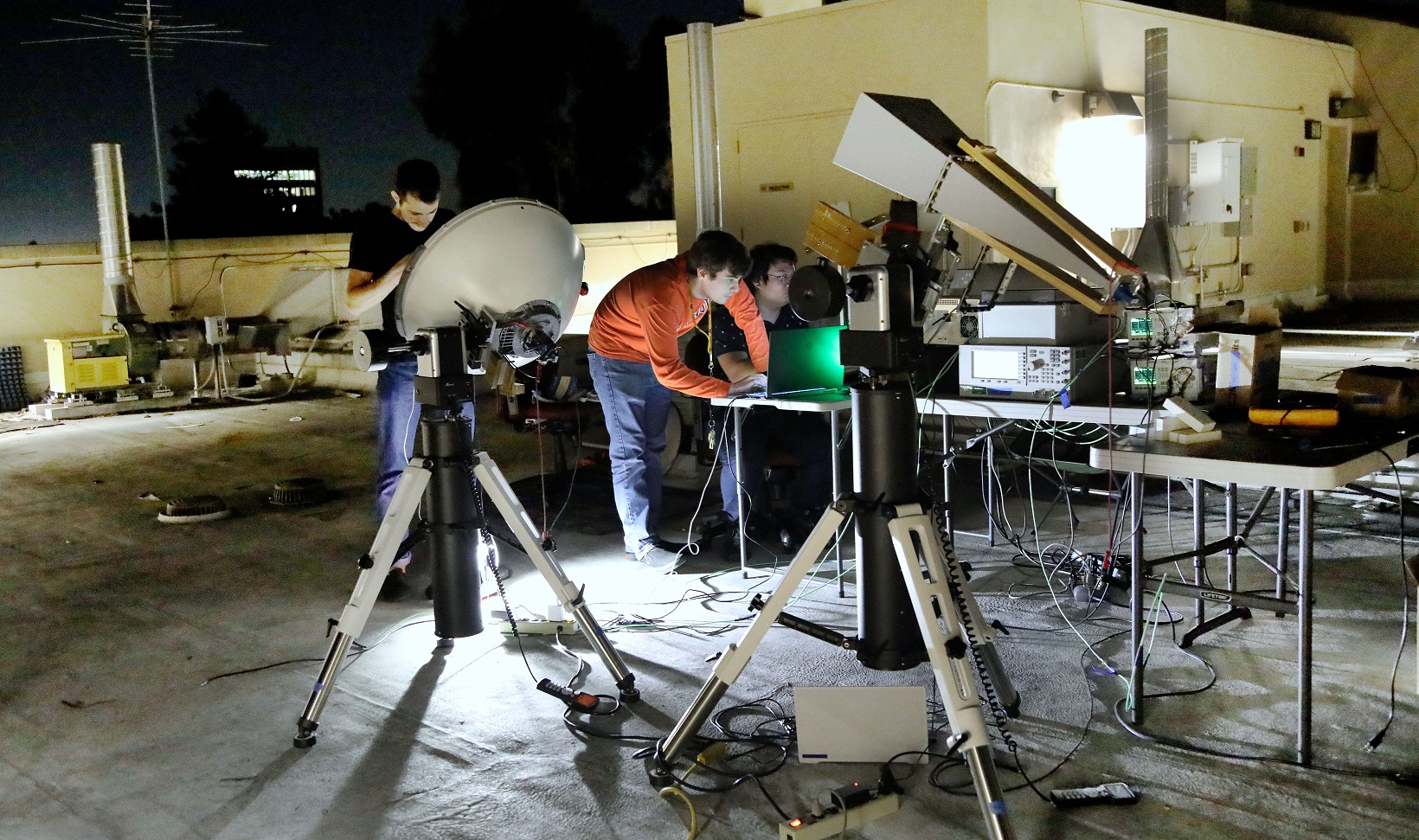}
   \caption{Full ground station assembled before the July 28th, 2023 UTC nighttime pass.}
   \label{fig:fullstation} 
\end{figure}

The orbit-tracking mount coordinated pointing at MAPLE using a two-line element (TLE) collected within 24 hours of the pass from the US Space Force website \cite{space-track}. Immediately before the pass, the mount's built-in GPS system was calibrated to known celestial objects (e.g., stars, planets, and/or the moon) through an optical sight aligned with the antenna's main beam. The full ground station, assembled before the beam-to-Earth attempt, is pictured in Fig.\ref{fig:fullstation}. 

\begin{figure*}[t]
\centering
   \includegraphics[width=0.85\textwidth]{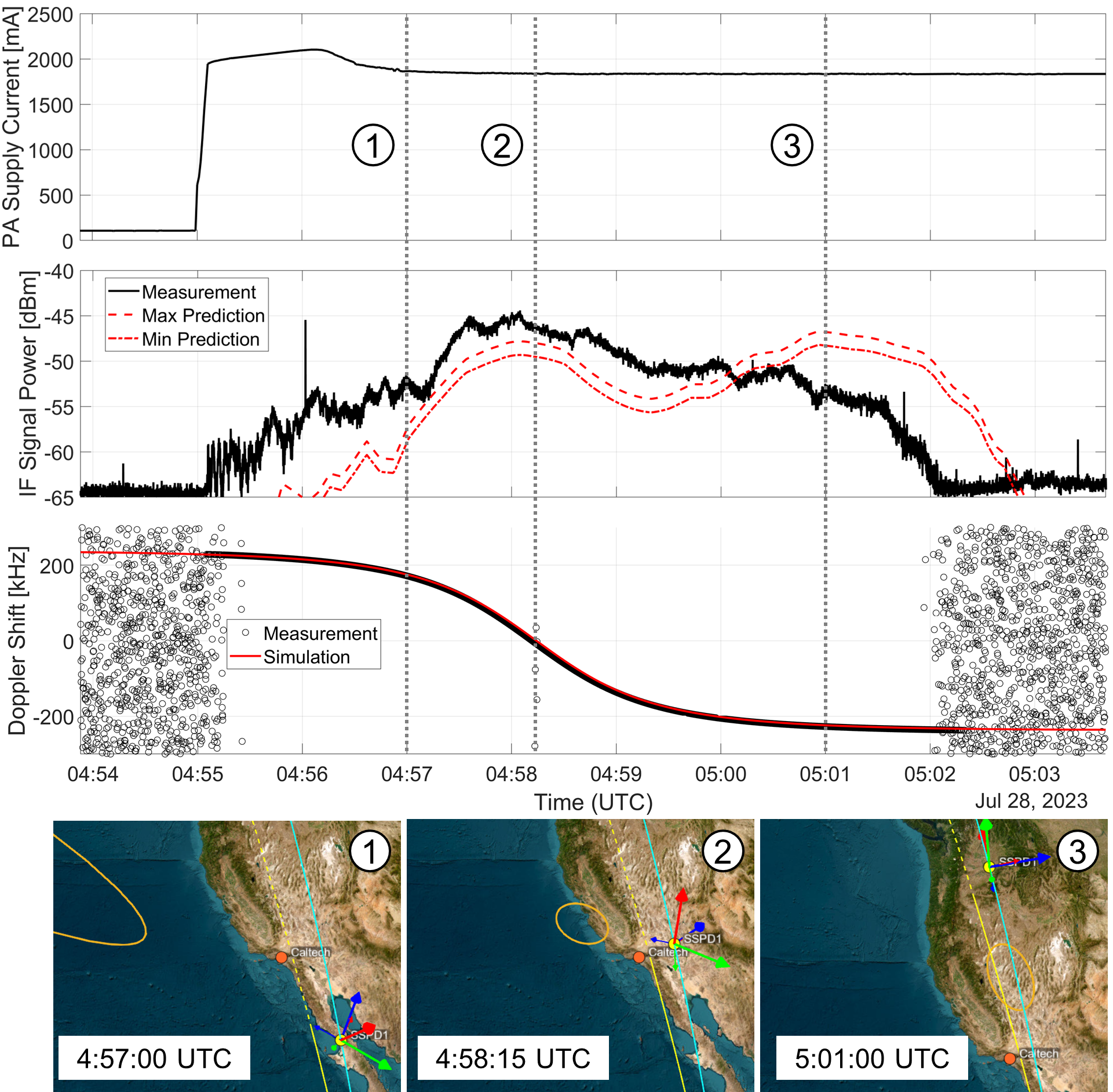}
   \caption{Beam-to-Earth experiment results from 7/28/2023 UTC. Top to bottom: power amplifier current draw for RFIC1, IF power received at the digitizer, Doppler shift from IF center frequency (5MHz), and orbital tracking maps showing MAPLE during the pass. Orange ellipses in the bottom row depict the -3dB isopower contour for MAPLE's main beam.}
   \label{fig:bte2} 
\end{figure*}

\subsection{Beam-to-Earth Results}
Three beam-to-Earth experiments were attempted on the following dates: May 23rd, 2023 UTC, June 30th, 2023 UTC, and July 28th, 2023 UTC. In all of the experiments, MAPLE's beam was successfully detected with the expected Doppler shift and power-level at a time consistent with the orbital predictions and ephemeris data from Vigoride-5. The final experiment on July 28th, 2023 UTC provided the largest continuous set of data; those results are presented in Fig. \ref{fig:bte2}. 

The power amplifier current, measured using flight computer telemetry, indicates the interval of time during which MAPLE was actively transmitting power: starting at 4:55 UTC and ending after MAPLE crosses the horizon. MAPLE's turn-on time corresponded to its crossing the horizon and matches the time at which IF power begins to exceed the system noise floor. MAPLE was closest to the ground station at 4:58:15 UTC; this is approximately the time of peak IF signal power and zero Doppler shift, as predicted.

Snapshots of MAPLE's location relative to Caltech and the -3dB isopower contour of its main beam are shown at the bottom of Fig. \ref{fig:bte2} for three different times. Unfortunately, the ground station was never within MAPLE's -3dB contour, due to the host's inability to maintain alignment, thus reducing peak IF power relative to the power budget prediction. Curves showing the range of predicted IF signal power in Fig. \ref{fig:bte2} account for pointing mismatch using a beam pattern measured on the ground with the EM. The exact pattern in orbit was likely slightly different due to the difference in environment and phase settings. This likely contributed to the discrepancy between predicted and measured IF signal power.

The signal was lost as MAPLE descended below the horizon at approximately 5:02 UTC. The measured Doppler shift over time closely matches a prediction made using the known payload orbital dynamics. Close alignment between IF signal power and power prediction, alignment between Doppler shift measurement and frequency prediction, and spacecraft telemetry confirm to a high degree of certainty that the signal detected was MAPLE.

\section{Anomalous Results}
``Amomalous" used herein describes any observable and measurable phenomenon in the data collected from the MAPLE Flight Model (FM) during in-orbit testing which was not present in the data collected from the MAPLE Engineering Model (EM) on Earth during normal operation. In most cases, these phenomenon were adverse. A number of these phenomenon are relevant to the discussion for the purpose of completeness, understanding the system better, and identifying potential areas for improvement. These phenomenon, called ``symptoms" below, are presented, followed by a discussion on the testing performed to uncover the possible cause(s).

\subsection{Symptoms}
\subsubsection{Rectified Power}

\begin{figure}[t]
\centering
\begin{subfigure}[b]{0.48\textwidth}
   \includegraphics[width=1\linewidth]{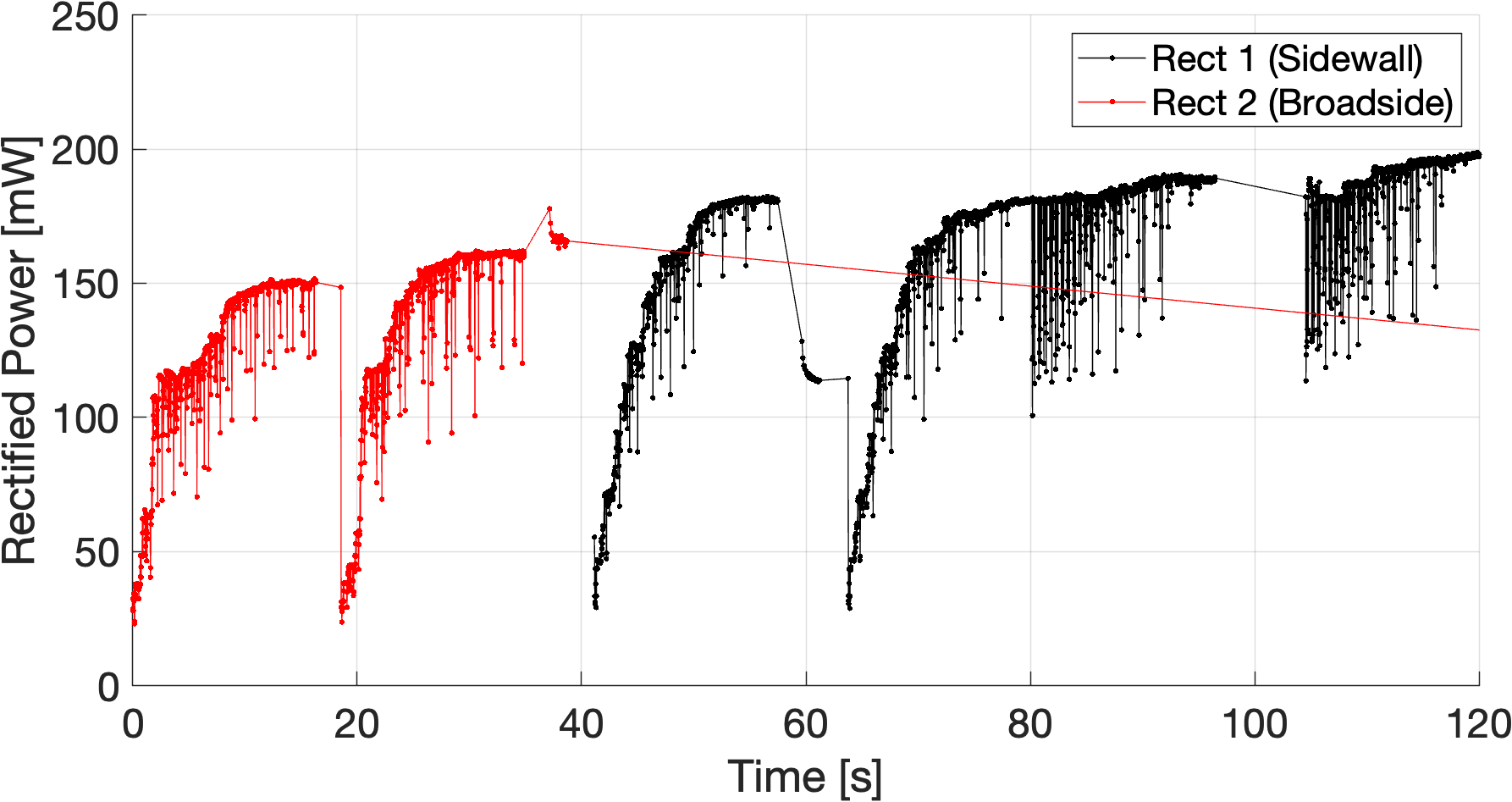}
   \caption{On-Earth test performed directly prior to launch}
   \label{fig:IOS_integration} 
\end{subfigure}

\begin{subfigure}[b]{0.48\textwidth}
   \includegraphics[width=1\linewidth]{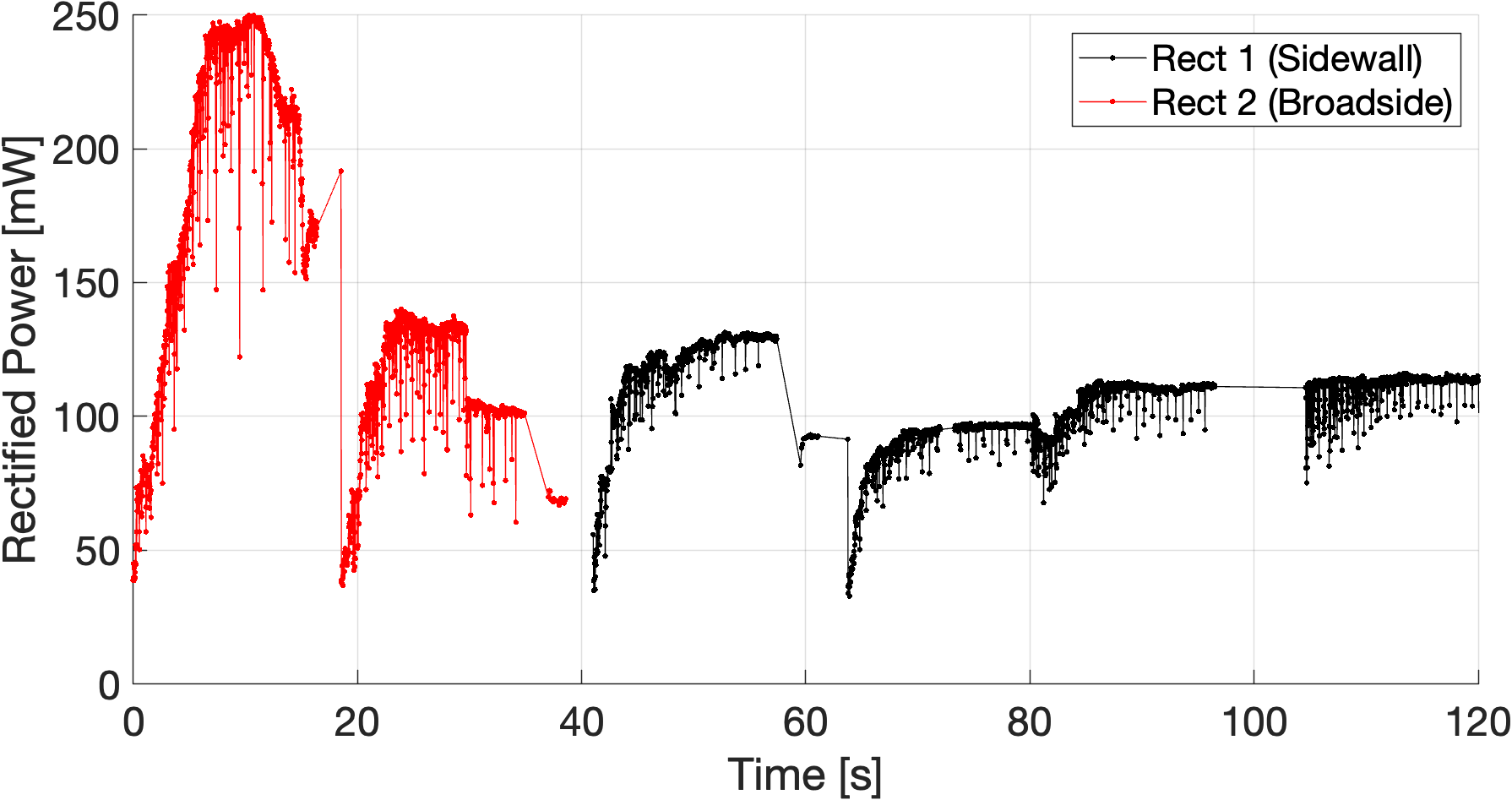}
   \caption{In-orbit test performed on March 3rd, 2023}
   \label{fig:IOS_MAT1}
\end{subfigure}

\caption{The first part of an ``Initial Operation Sequence" (IOS): Two optimizations to Rectenna 2, phase recall to Rectenna 2, one optimization to Rectenna 1, phase recall to Rectenna 1, and then three successive optimizations to Rectenna 1, each optimization starting its search at the last setting of the previous optimization.}
\end{figure}

Primary among the anomalous symptoms observed during in-orbit experiments are properties of the rectified power waveforms produced at the rectenna. During optimizations, rectified power should increase more or less monotonically, save for expected and occasional dramatically low dips in power as the algorithm stumbles on local maxima in the search space. After 10-30 seconds, the algorithm should stabilize at a maximum power (often close to the global maximum), which it is able to hold for the remainder of the optimization period. Inability to remain at this maximum, demonstrated by a slow drift downward (``drooping") or sporadic changes to the output power, indicates the system is varying in time and undermining the previously computed search space optimum.

Some of the results measured in-orbit are anomalous with respect to both the absolute power level achieved by normal optimization runs and the shape of the rectified power curve. With respect to the absolute power level: after the first optimization, the system is unable to reproduce peak power at the receiver. Optimizations stabilize, instead, at lower power values.

With respect to the shape of the power curves: optimizations often failed to produce the typical increasing asymptotic curve that is indicative of a healthy, working system. Instead, curves showed sporadic behavior including dramatic but continuous decreases in rectified power and disjointed decreases in rectified power. This is evidenced by the data presented from the first in-orbit test performed on March 3rd, 2023 in Fig. \ref{fig:IOS_MAT1}. 

Presented in Fig. \ref{fig:IOS_integration} is data collected during integration testing wherein the expected power curves are produced by optimization. During a typical optimization, power levels repeatedly and consistently asymptotically increase and then stabilize at some maximum. When an optimization begins at the stable point of the previous optimization, as is done in the second half of the data in Fig. \ref{fig:IOS_integration}, there should be marginal increases in rectified power, but generally a flat curve.

However, in Fig. \ref{fig:IOS_MAT1}, a different pattern emerges. The first optimization, to rectenna 2, initially rises to 250mW but, instead of settling, begins to drop continuously, finishing at 175mW. The second optimization, also to rectenna 2, does not approach the previous peak of 250mW, but momentarily stabilizes at 140mW before dropping to 100mW. The third optimization, to rectenna 1, has the appropriate shape, but peaks at 130mW, lower than the integration peak of 175mW. The fourth optimization peaks still lower, but demonstrates the expected flat curves. The last two optimizations are similarly low power but flat.

Optimizations performed while PAs drew less power demonstrated lower performance degradation. Power reduction was accomplished by either lowering output PA gain directly or by only using a subset of the array elements instead of all 32. In the latter case, optimizations had nominal shapes and peak power levels given the number of active elements.

\subsubsection{PA Supply Current}
\begin{figure*}[t]
\centering
   \includegraphics[width=1\linewidth]{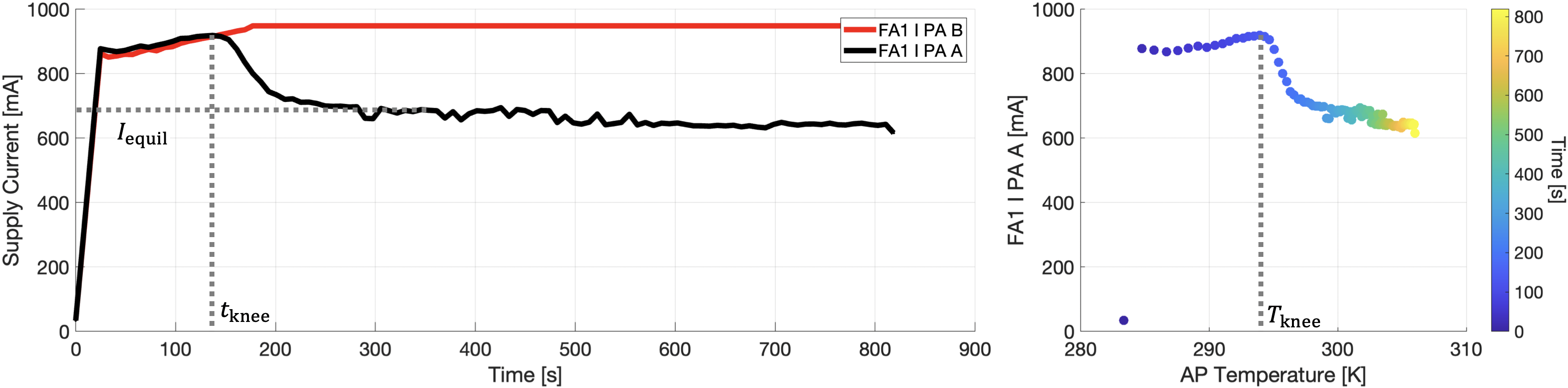}
   \caption{Supply current drop anomaly on DB1. Left: Supply current for RFIC1 (lines A and B) measured over time. Right: Supply current plotted against board temperature. $t_\text{knee}$, $T_\text{knee}$, and $I_\text{equil}$ labeled.}
   \label{fig:Tknee} 
\end{figure*}

Power amplifier (PA) supply currents measured during in-orbit tests also appeared in some cases. PA supply lines are routed from the flight computer (FC) to the AP where they are voltage-regulated and then passed to the FA for routing to the two RFICs. Four PA supply lines (1A, 1B, 2A, 2B) route to the two RFICs (1/2); for RFIC1, quads 1 and 2 share supply line 1A and quads 3 and 4 share supply line 1B. On the AP, these lines are voltage- and current-monitored during regular ``health checks." These health checks provide the telemetry data referenced herein which are indicative of PA operation during testing.

Health checks measured supply lines 1A and 2A slowly rising ($\approx 5\%$) and then rapidly dropping ($\approx 24\%$) after PAs were on for a sufficiently long time, as shown in Fig. \ref{fig:Tknee}. This point in time, called $t_\text{knee}$, varied between RFICs ($t_\text{knee}$, RFIC1 $\approx$ 100s, $t_\text{knee}$, RFIC2 $\approx$ 20s) and between tests. For reasons discussed below, it's likely the drop in current is triggered by increased temperature; the temperature of the AP at $t_\text{knee}$ is called $T_\text{knee}$. Both quantities are labelled in Fig. \ref{fig:Tknee}.

PA supply lines 1B and 2B never dropped - they were observed rising about 10\% and then stabilizing, even for very long tests.

\subsubsection{Chip-Chip Interference Pattern}
MAPLE is equipped with algorithms that create interference patterns between the two RFICs as a way to characterize chip-to-chip coherence. To accomplish this, the array of phases for one RFIC is held constant while the array of phases for the other RFIC is progressively and discretely swept through a list of phases from $0^\circ$ to $360^\circ$. To add, for example, 10$^\circ$ to one chip, each core's previously measured CMU phase function is inverted to solve for the necessary unique phase code that will accomplish this phase shift. On Earth, these algorithms worked, successfully producing interference patterns with periods commensurate with the programmed phase step.

\begin{figure}[t]
\centering
\begin{subfigure}[b]{0.48\textwidth}
   \includegraphics[width=1\linewidth]{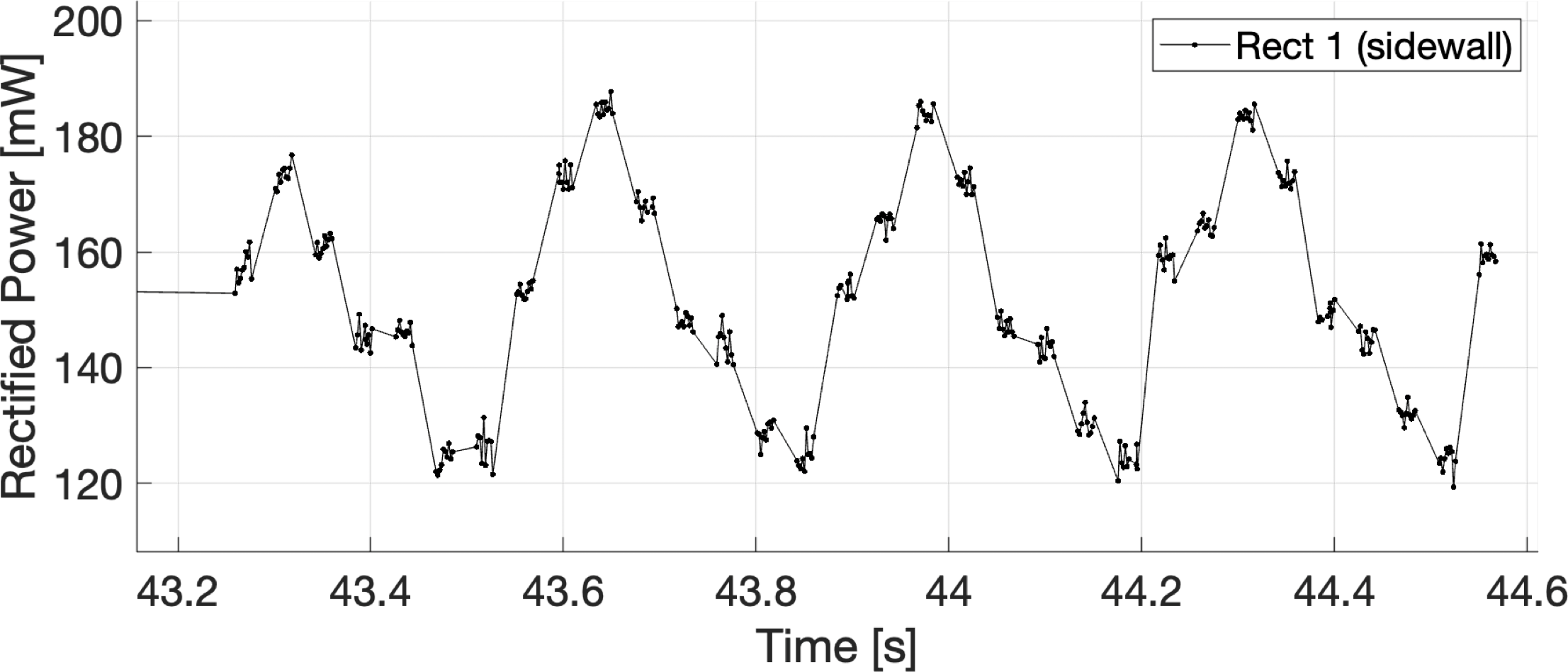}
   \caption{On-Earth test performed directly prior to launch}
   \label{fig:inteference_integration} 
\end{subfigure}

\begin{subfigure}[b]{0.48\textwidth}
   \includegraphics[width=1\linewidth]{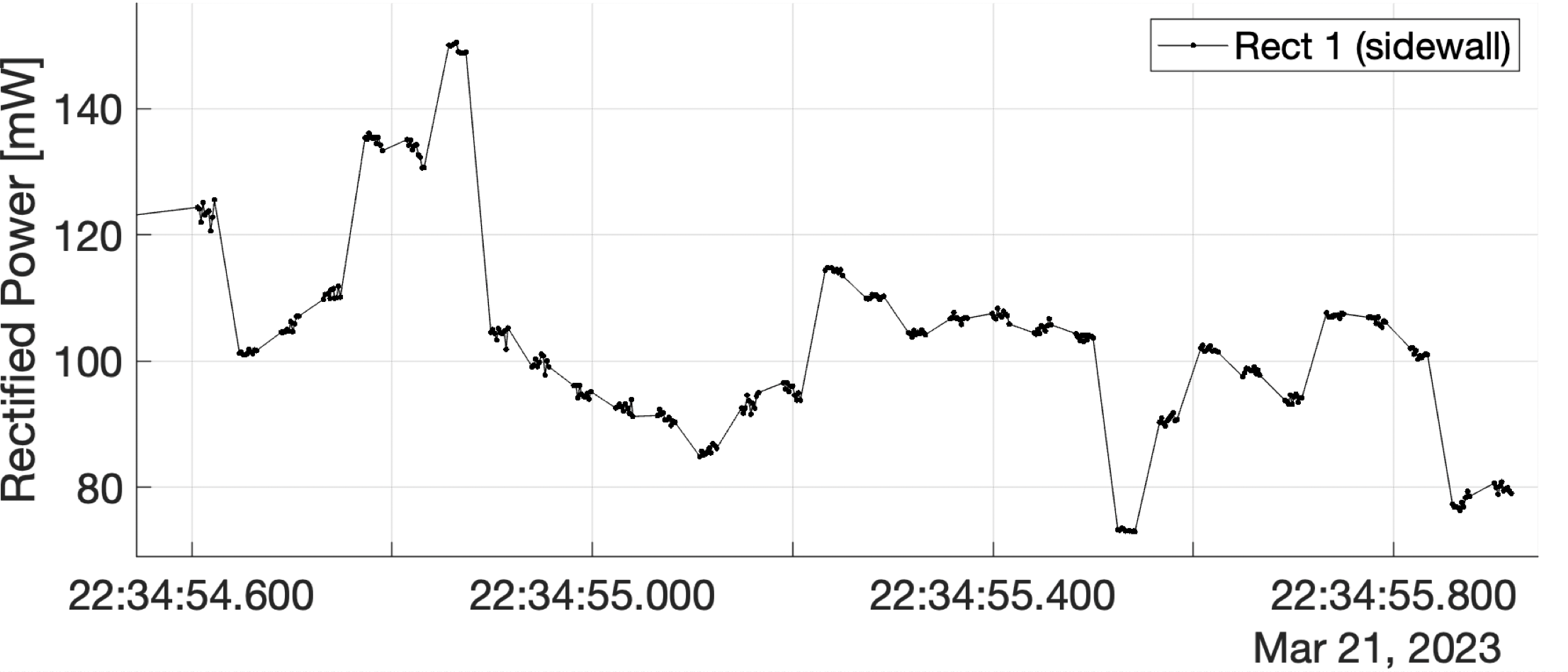}
   \caption{In-orbit test performed on March 21, 2023}
   \label{fig:inteference_MAT6}
\end{subfigure}

\caption{Rectified power during linear sweep of phase offset added to DB2.}
\label{fig:inteference_anomaly}
\end{figure}

In-orbit, the same algorithm did not produce clean cyclic interference patterns, thus undermining at least one of the embedded assumptions. When run, the algorithm produces less predictable interference patterns, as shown in Fig. \ref{fig:inteference_anomaly}. Data from both tests was taken $\approx 43$s after turning the PAs on: a moderate ``warm-up" period but not long enough to let the system ``settle" completely. This points to significant changes to the CMU phase function(s).

\subsubsection{Degradation Metrics Over Time}
These anomalies, and system performance in general, are captured by parameters like $t_\text{knee}$, the time between system turn-on and the PA supply current drop, $T_\text{knee}$, the temperature of the AP when PA supply current drops\footnote{As demonstrated in Fig. \ref{fig:Tknee}, plotting PA current against temperature, it appears that PA current drops in response to the crossing of some threshold. This threshold is called $T_\text{knee}$.}, and $I_\text{eq}$, the supply current after a sufficiently long time (seven minutes).

Low values for $t_\text{knee}$ and $T_\text{knee}$ indicate the RFIC's absolute temperature rise or {\it sensitivity} to temperature rise; both could be indications of system degradation. Likewise, low $I_\text{eq}$ could serve as an indication of system degradation.

\begin{figure}[t]
\centering
   \includegraphics[width=\linewidth]{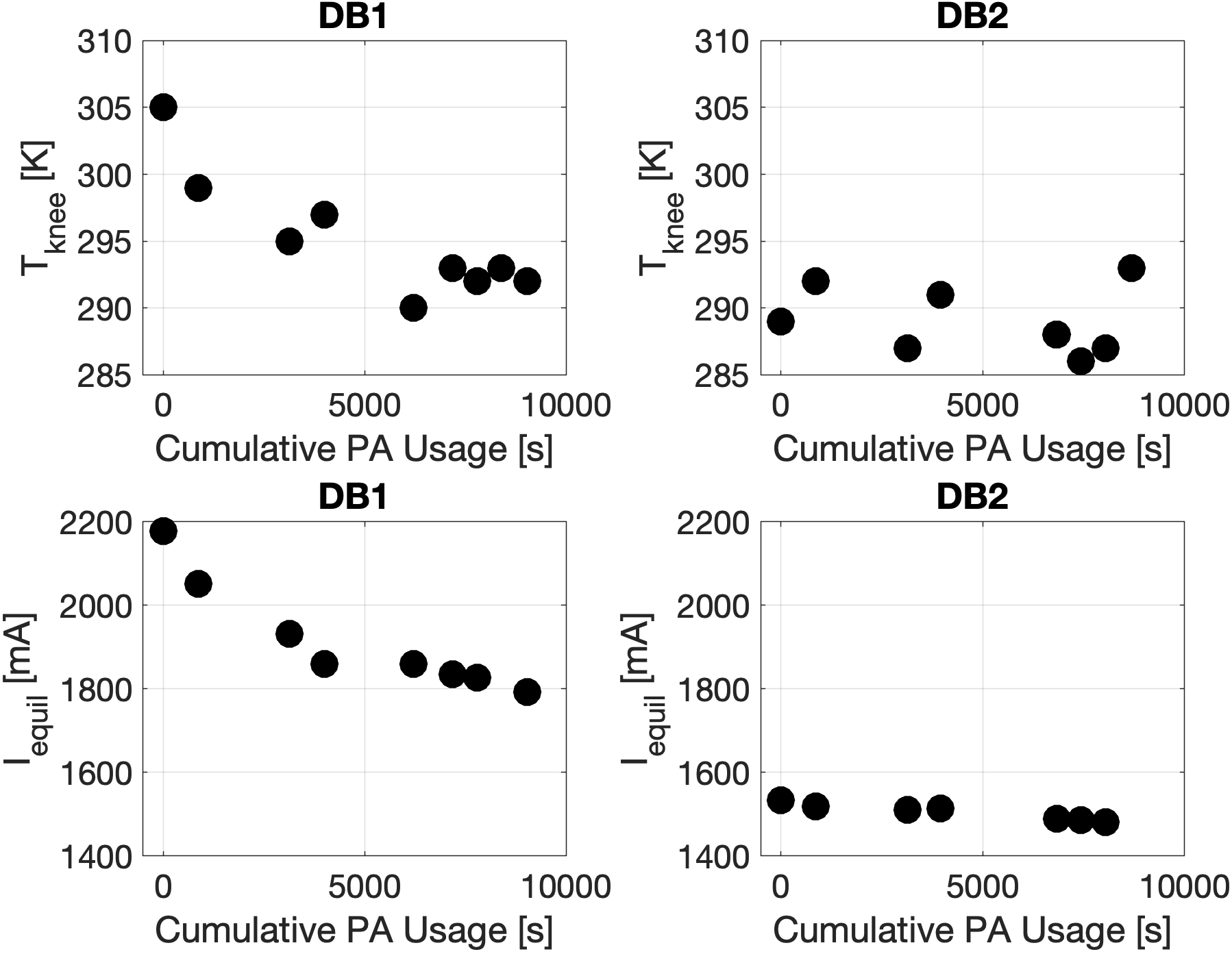}
   \caption{Long-term system degradation as indicated by two metrics, $T_\text{knee}$ and $I_\text{eq}$}
   \label{fig:degradation} 
\end{figure}

These metrics were tracked during simple, long-duration tests performed over the mission lifetime to characterize long-term system degradation. In Fig. \ref{fig:degradation}, it's clear RFIC1 $T_\text{knee}$ and $I_\text{equil}$ drop over time, with cumulative PA usage. RFIC2 appears much more sensitive to abnormalities, with the supply current dropping immediately upon being turned on. Thus, $T_\text{knee}$ is the nominal board temperature when the system is turned on; for this reason, we see no long-term degradation on RFIC2.

Thermal metrics and results presented in Fig. \ref{fig:before_after} demonstrate different levels of degradation:
\begin{enumerate}
    \item Short-term, reversible output power degradation is the primary degradation, observed in the mission's first tests and responsible for significant output power decreases during testing. 
    \item Long-term degradation, observed in the decline of thermal metrics over time (Fig. \ref{fig:degradation}), exacerbates output power decreases and appears to worsen over time. Additionally, an $\approx$23\% decrease in peak achievable rectified power over eight months (Fig. \ref{fig:before_after}) quantifies long-term power output degradation.
    \item Individual, core-level degradation was observed in lower output amplitudes in six cores in Fig. \ref{fig:before_after}.
\end{enumerate}
Symptom \#3 (the six degraded cores) may be the driving factor in Symptom \#2 (long-term degradation.) Additionally, Symptoms \#2 (long-term degradation in thermal metrics and output power) and \#3 (individual core damage) may be the result of normal flight-related ageing, or the prolonged effects of the issue(s) causing Symptom \#1.

\subsection{Anomalies: Likely Cause}
Anomalies likely stem from a single issue: poor thermal contact between the RFIC and the heatsink. Though designed to operate at slightly above chassis temperature, the RFICs, due to damage to the thermal bond with the aluminum heatsink pedestal, progressively heat up when PAs are turned on. The rate and asymptotic chip temperature are functions of thermal bond degradation---RFIC2 might perform generally worse because its bond is degraded further. This means that RFIC2 crosses thermal thresholds faster and stabilizes at a higher temperature overall.

As the temperature of the RFICs rise, PA output amplitudes and CMU phase functions vary, thus introducing instability to the optimization. Though only select elements' phases are digitally varied at once, the other elements' output phasors vary as the chip overheats. This produces erratic and non-monotonic optimization curves. Final rectified power values are lower even after the RFICs stabilize as a result of temperature-induced, degraded PA performance.

Nominal performance when only a subset of the quads is active, or an overall lower PA output power is programmed, is the result of lower power dissipation and, thus, lower steady-state temperature. Transitory temperature-induced performance degradation is nonlinear with temperature and, thus, degradation is only observed under full- or mostly full-power operation.
Performance restores when the chips cool down; turning the PAs off, pausing briefly, and then restarting restores full functionality until the chips reheat.

Inability to generate a sinusoidal interference pattern between chips is indeed the result of temperature-induced variations on the CMU phase functions. The functions at higher temperatures are different, and the underlying assumptions about what codes generate what phase offsets is broken. Time variations in core output also introduce aperiodicity.

A temperature threshold, $T_\text{knee}$, crossed also induces PA current supply drop, but only on supply line A. Supply line B experiences an increase in supply current draw as a result of temperature-dependent current draw to the underlying PA transistor.

This hypothesis is the result of extensive testing and verification, in space and in TVAC chambers on Earth using the EM. When thermistors were mounted {\it close} to the chip\footnote{The RFICs are thermally heat sunk on one side and flip-chipped to an interposer installed on the backside of the flexible PCB on the other. The closest location, thermally, to the chip is on the front side of the flexible PCB, opposite the chip. This is where a thermistor was installed. During proper heatsink operation, the temperature difference between the RFIC and the thermistor is only 10K.}, in an EM model of MAPLE, during on-Earth TVAC testing, temperatures in excess of 330K and 360K were measured for RFIC1 and RFIC2, respectively. RFIC1, at this temperature, behaved nominally while RFIC2 presented all of the aforementioned symptoms. All testing indicates that the fact that anomalous behavior presented on RFIC2 in both space {\it and} TVAC testing, on two different MAPLE models, is probably coincidental. RFICs 1 and 2 are constructed and wired almost identically and are arranged symmetrically vis-a-vis thermal pathways.

The discrepancy between chips is likely explained by discrepancies in thermal bond damage. When the EM was placed in a thermal chamber under ambient pressure, symptoms were replicated on both RFICs only when air temperature was in excess of 415K. This temperature is close to the limit of device models available from the IC foundry and likely high enough to induce the symptoms observed. An experiment on the ground where the RFIC was operated under ambient conditions with its heatsink intentionally removed showed similar discrepancies compared with a chip whose heatsink was intact. 

A difference in the temperature necessary to replicate symptoms in TVAC and in thermal chamber (ambient pressure) can be explained by poor RFIC-epoxy bond quality. When the intended thermal pathway is compromised, heat flows through the flex array, which has higher thermal resistance and increases the temperature of the RFIC. This effect is more pronounced in vacuum due to the absence of convective heat transfer.

The RFIC-epoxy bond can be degraded both by vibration-induced failure and by workmanship flaws not revealed during environmental testing. Additionally, ground testing indicates contact pressure can affect the integrity of the bond, as expected based on the discussion in \cite{gluck2002mountings}. MAPLE's design does not apply pressure to the RFIC-epoxy bond, potentially increasing sensitivity to workmanship flaws and resulting in degraded performance. Ground-testing to further understand the failure-mode, in order to inform future missions, is ongoing.  

\section{Conclusion \& Future Work}
This paper discussed the results of a ten month, LEO mission that is, to the best of our knowledge, the first documented demonstration of wireless energy transfer in space. Further, this power was transferred using flexible, lightweight arrays driven by custom RFICs. The design of the system and its components, including discussion of the integrated circuit, custom co-cured popup dipole antennas, flexible electronics, focusing algorithms, thermal-mechanical design, and space qualification testing were presented. Results for eight months of in-orbit testing were discussed, including demonstrations of focusing and WPT in space, rapid refocusing between different targets, and individual radiator core performance. Long-term system performance studies and a comparison to on-Earth performance were presented as well.

Moreover, presented above are results from the on-Earth detection of wireless power transferred from a payload in orbit. Despite its small size and output power, the flexible aperture could focus power successfully to deliver power up to 20dB above the ground station noise floor and with the expected Doppler shift signature.

The results from this mission were highly informative and identified a number of key areas for improvement: namely, the system's thermal design. Likewise, mission successes serve as a proof-of-concept and a step forward for the practical implementation of SSP while system weaknesses offer areas to build upon in future projects.

Future project directions include development of sinkless, radiatively cooled, flexible wireless power arrays; integration of novel functionality such as shape sensing \cite{MIZRAHI_FLEXIBLE_RECONSTRUCTION, FIKES_FRAMEWORK_MTT} and/or dynamic impedance matching \cite{ayling_MPO}; larger-scale systems; novel time-synchronization schemes; and integration with flexible photovoltaic systems. 

These results and developments promise to bring us one step closer to the vision of clean, affordable energy provided by arrays in space.

\section{Acknowledgements}
The authors would like to thank R. Madonna, S. Pellegrino, H. Atwater, C. Sommer, E. Gdoutos, A. Truong, F. Wiesem\"{u}ller, K. Ubamanyu, A. Cortez, M. Manohara, J. Pederson, M. Kelzenberg, A. Wen, J. Sauder, J. Rivera, H. Lim, E. Suh, W. Chun, A. Safaripour, M. Howard, H. Lambert, K. Spencer, A. Daniel, T. Roper, D. Johnson, and L. Day for their support, contributions, and discussions.

\bibliographystyle{elsarticle-num}
\bibliography{main_bib}

\end{document}